\documentclass[twocolumn,times]{aastex63}

\received{XXXX XX, 2022}
\shorttitle{X-ray Unveiling Events}
\shortauthors{Yu et al.}

\usepackage{graphicx}
\usepackage{multirow}
\usepackage{color}
\usepackage[figuresright]{rotating}
\usepackage{longtable}
\usepackage{mathrsfs,amsmath}
\usepackage{epstopdf}
\newsavebox{\tablebox}
\usepackage{hyperref}
\usepackage{times}
\usepackage{longtable}
\usepackage{array}
\usepackage{booktabs}

\newcommand{\kms}{\ifmmode {\rm km\ s}^{-1} \else km s$^{-1}$\ \fi}
\newcommand{\ergs}{\ifmmode {\rm erg\ s}^{-1} \else erg s$^{-1}$ \fi}
\newcommand{\lb}{\ifmmode L_{\rm bol} \else $L_{\rm bol}$\fi}
\newcommand{\mbh}{\ifmmode M_{\rm BH}  \else $M_{\rm BH}$\ \fi}
\newcommand{\msun}{M_{\odot}}
\newcommand{\nh}{\ifmmode N_{\rm H} \else $N_{\rm H}$\ \fi}
\newcommand{\pb}{\ifmmode P_{\rm B}  \else $P_{\rm B}$\ \fi}
\newcommand{\chandra}{{\it Chandra\/}}

\newcommand{\xray}{\hbox{X-ray}}
\newcommand{\xid}{\hbox{XID 403}}
\newcommand{\rband}{\hbox{$R$-band}}
\newcommand{\al}{\hbox{$\alpha_{\rm OX}\textrm{--}L_{\rm 2500~{\textup{\AA}}}$}}

\begin{document}
\title{X-ray Unveiling Events in a $z \approx 1.6$ Active Galactic Nucleus in the 7 Ms {{\em Chandra}} Deep Field-South}

\author{Li-Ming~Yu}
\affiliation{School of Astronomy and Space Science, Nanjing University, Nanjing, Jiangsu 210093, China}
\affiliation{Key Laboratory of Modern Astronomy and Astrophysics (Nanjing University), Ministry of Education, Nanjing 210093, China}

\author{Bin~Luo}
\affiliation{School of Astronomy and Space Science, Nanjing University, Nanjing, Jiangsu 210093, China}
\affiliation{Key Laboratory of Modern Astronomy and Astrophysics (Nanjing University), Ministry of Education, Nanjing 210093, China}

\author{W.~N.~Brandt}
\affiliation{Department of Astronomy \& Astrophysics, 525 Davey Lab, The Pennsylvania State University, University Park, PA 16802, USA} 
\affiliation{Institute for Gravitation and the Cosmos, The Pennsylvania State University, University Park, PA 16802, USA}
\affiliation{Department of Physics, 104 Davey Lab, The Pennsylvania State University, University Park, PA 16802, USA}

\author[0000-0002-8686-8737]{Franz E. Bauer}
\affiliation{Instituto de Astrof{\'{\i}}sica and Centro de Astroingenier{\'{\i}}a, Facultad 
de F{\'{i}}sica, Pontificia Universidad Cat{\'{o}}lica de Chile, Casilla 306, 
Santiago 22, Chile} 
\affiliation{Millennium Institute of Astrophysics (MAS), Nuncio Monse{\~{n}}or 
S{\'{o}}tero Sanz 100, Providencia, Santiago, Chile} 
\affiliation{Space Science Institute, 4750 Walnut Street, Suite 205, Boulder, 
CO 80301, USA}

\author{D.~De Cicco}
%\affiliation{Instituto de Astrofísica, Pontificia Universidad Católica de Chile, Av. Vicuña Mackenna 4860, 7820436 Macul, Santiago, Chile}
\affiliation{Instituto de Astrof\'{i}sica, Pontificia Universidad Cat\'{o}lica de Chile, Av. Vicu\~{n}a    Mackenna 4860, 7820436 Macul, Santiago, Chile}
\affiliation{Millennium Institute of Astrophysics (MAS), Nuncio Monse\~nor Sotero Sanz 100, Of. 104, Providencia, Santiago, Chile}
\affiliation{Dipartimento di Fisica, Università degli Studi di Napoli ``Federico II'', via Cinthia 9, 80126 Napoli, Italy}

\author{A.~Fabian}
\affiliation{Institute of Astronomy, Madingley Road, Cambridge CB3 0HA, UK}

\author{R.~Gilli}
\affiliation{
INAF – Osservatorio di Astrofisica e Scienza dello Spazio di Bologna, Via P. Gobetti 93/3, 40129 Bologna, Italy}

\author{A.~Koekemoer}
\affiliation{Space Telescope Science Institute, 3700 San Martin Drive,
Baltimore, MD 21218, USA}

\author{M.~Paolillo}
\affiliation{Dip.di Fisica Ettore Pancini, University of Naples
``Federico II'', C.U. Monte S. Angelo, Via Cintia, 80126, Naples, Italy}
\affiliation{INFN -- Sez.di Napoli, Via Cintia, 80126, Naples, Italy}
\affiliation{INAF -- Osservatorio Astronomico di Capodimonte, salita Moiariello 16, 80131 Napoli, Italy }

\author{D.~P.~Schneider}
\affiliation{Department of Astronomy \& Astrophysics, 525 Davey Lab, The Pennsylvania State University, University Park, PA 16802, USA} 
\affiliation{Institute for Gravitation and the Cosmos, The Pennsylvania State University, University Park, PA 16802, USA}

\author{O.~Shemmer}
\affiliation{Department of Physics, University of North Texas,
Denton, TX 76203, USA}

\author{P.~Tozzi}
\affiliation{INAF -- Osservatorio Astrofisico di Arcetri, Largo
E. Fermi 5, I-50125, Florence, Italy}

\author{Jonathan~R.~Trump}
\affiliation{Department of Physics, University of Connecticut, Storrs, CT 06269, USA}

\author{C.~Vignali}
\affiliation{Dipartimento di Fisica e Astronomia ``Augusto Righi'', Universit\`{a} degli Studi di Bologna, Via Gobetti 93/2, I-40129 Bologna, Italy}
\affiliation{INAF -- Osservatorio di Astrofisica e Scienza dello Spazio di Bologna, Via Gobetti 93/3, I-40129 Bologna, Italy}

\author{F.~Vito}
\affiliation{INAF -- Osservatorio di Astrofisica e Scienza dello Spazio di Bologna, Via Gobetti 93/3, I-40129 Bologna, Italy}
\affiliation{Scuola Normale Superiore, Piazza dei Cavalieri 7, I-56126 Pisa, Italy}

\author{J.-X.~Wang}
\affiliation{CAS Key Laboratory for Research in
Galaxies and Cosmology, Department of Astronomy,
University of Science and Technology of China, Hefei, Anhui 230026, China}
\affiliation{School of Astronomy and Space Sciences, University of Science and Technology of China, Hefei 230026, China}

\author{Y.~Q.~Xue}
\affiliation{CAS Key Laboratory for Research in
Galaxies and Cosmology, Department of Astronomy,
University of Science and Technology of China, Hefei, Anhui 230026, China}
\affiliation{School of Astronomy and Space Sciences, University of Science and Technology of China, Hefei 230026, China}

\begin{abstract}
We investigate the extreme \xray\ variability of a $z = 1.608$ active galactic nucleus in the 7~Ms \chandra\ Deep Field-South (XID 403), which showed two significant \xray\ brightening events. In the first event, XID~403 brightened by a factor of $> 2.5$ in $\lesssim6.1$ 
rest-frame 
days in the observed-frame 0.5--5 keV band. The event lasted for $\approx5.0$--7.3 days, and then XID 403 dimmed by 
a factor of $>6.0$ in $\lesssim6.1$ days. After $\approx1.1$--2.5 years
in the rest frame (including long observational gaps), it brightened again with the 0.5--5~keV flux increasing by
a factor of $>12.6$. The second event 
lasted over 251 days and the source remained bright until
the end of the 7~Ms exposure. The spectrum is a steep power law (photon index $\Gamma=2.8\pm0.3$) without obscuration during 
the second outburst, and the rest-frame 2--10 keV luminosity reaches  $1.5^{+0.8}_{-0.5} \times 10^{43}~\rm erg~s^{-1}$;
there is no significant spectral evolution within this epoch.
The infrared-to-UV spectral energy distribution of \xid\ is dominated
by the host galaxy.
There is no significant optical/UV variability and 
$R$-band (rest-frame $\approx2500$~\AA) 
brightening contemporaneous with the \xray\ brightening.
The extreme X-ray variability 
is likely due to two X-ray unveiling events, where
the line of sight to the corona is no longer shielded by 
high-density gas clumps in a small-scale dust-free absorber. \xid\ is probably  a high-redshift 
analog of local narrow-line Seyfert 1
galaxies, and the X-ray absorber is  a 
powerful accretion-disk wind. 
%with variable covering factor  and column density that is driven by super-Eddington accretion of the central black hole.  
On the other hand, we cannot exclude the possibility that \xid\ is an unusual candidate for tidal disruption events.

\end{abstract}


\section{Introduction} \label{sec:intro}

X-ray emission appears  ubiquitous from  active galactic nuclei (AGNs), and it is considered largely to originate from the accretion-disk corona in the vicinity of the central supermassive black hole (SMBH). \xray\ photons are produced from inverse-Compton scattering of the optical and ultraviolet (UV) \hbox{accretion-disk} photons by the coronal relativistic electrons \citep[e.g.,][]{Liang1977,Sunyaev1980, Haardt1991,Done2010,Gilfanov2014,Fabian2017}.
Observations find that the \xray\ radiation of radio-quiet AGNs is related to their optical/UV radiation, and the relation is typically expressed as a negative correlation between  the \hbox{optical-to-X-ray}
{power-law} slope parameter  ($\alpha_{\rm OX}$)\footnote{$\alpha_{\rm OX}= 0.384~{\rm \log}(f_{\rm 2~{\textup{keV}}}/f_{\rm 2500~{\textup{\AA}}})$, where $f_{\rm 2~{\textup{keV}}}$ and $f_{\rm 2500~{\textup{\AA}}}$ are the \hbox{rest-frame} 2 keV and 2500 \AA\ flux densities.}
and $2500~{\textup{\AA}}$ monochromatic luminosity ($L_{\rm 2500~{\textup{\AA}}}$) across a broad range of AGN luminosities \citep[e.g.,][]{Steffen2006,Just2007,Grupe2010,Lusso2016,Pu2020,Timlin2020}. 

%\hbox{$\alpha_{\rm OX}$--$L_{\rm 2500~{\textup{\AA}}}$} relation  
%Given the ubiquitous \xray\ emissions from AGNs, \xray\ surveys have been utilized to classify AGNs. It provides a good census of cosmic accreting SMBHs \citep[e.g.,][]{BA2015}. Observed \xray\ emissions from the most of the  \hbox{deep-survey} AGNs are not intrinsic  \citep[e.g.,][]{Luo2017}. They might be modified by absorption from gas or dusty torus \citep[e.g.,][]{Krolik1986,Levenson2002,Eguchi2009,Netzer2015}.
%Strong X-ray variability arise from galactic nuclei is usually considered to be attributed to the activities of the central supermassive black hole (SMBH).
X-ray variability is also a characteristic property of AGNs.
Observations of \xray\ variability in large samples of AGNs have revealed that the typical long-term \xray\ variability amplitude  is $\approx 20\textrm{--}50\%$ \citep[e.g.,][]{Grupe2001,Paolillo2004,Vagnetti2011,Gibson2012,Yang2016,Falocco2017,Maughan2019} with significant dependencies upon luminosity and other factors. Such \xray\ variability is generally attributed to  disk/corona instabilities or small fluctuations of the SMBH accretion rate.
Strong \xray\ variability events with flux varying by factors of $\gtrsim 2$ are rare \citep[][]{Yang2016,Timlin2020}, and they require additional mechanisms for interpretation.  Possible scenarios include
%In the inactive galaxy, extreme \xray\ variability  with \xray\ luminosities of $ \approx 10^{42}~\textrm{--}~10^{45}\rm~erg~s^{-1}$ are usually explained as tidal disruption events (TDEs; e.g., \citealt{Komossa2015,Saxton2021}).
%In general,  strong AGN \xray\ variability is explained by one of three scenarios: 
change of accretion rate \citep[e.g.,][]{LaMassa2015,MacLeod2016,MacLeod2019}, change of obscuration \citep[e.g.,][]{Matt03,Ni2020}, %disk reflection \citep[e.g.,][]{Ross2005}, 
and tidal disruption events (TDEs; e.g., \citealt{Gezari2021}), and they correspond to three different types of phenomena, i.e., substantial intrinsic variability of the disk/corona, variability of an external \xray\ absorber, and transient events.

AGN  radiation strength is directly linked to the accretion rate. Thus a large change of the accretion rate naturally results in strong optical/UV and \xray\ continuum variability. Such variability is often accompanied by broad emission-line variability (e.g., $\rm H\beta$) and the \xray\ spectrum generally shows no signs of obscuration.
%AGN with significant changes of accretion rate are usually identified based on changes in their optical/UV continuum and \hbox{broad-line} emission \citep[e.g.,][]{LaMassa2015,Parker2016,Parker2019}.
%AGN bolometric luminosity (\lb) is directly related to the accretion rate. The mechanism of change of accretion rate would result in change of the radiation from the accretion disk, thus producing the strong \xray\ variability. Changes of accretion rate are usually identified from changes of optical/UV continuum and broad emission lines. 
%Some of these AGNs have contemporaneous \xray\ observations, and they generally show corresponding \xray\ flux variability  with no apparent signatures of  \xray\ obscuration \citep[e.g.,][]{LaMassa2015,Parker2016,Parker2019}. 
An example is  \hbox{SDSS J015957.64$+$003310.5} reported by \cite{LaMassa2015}, which displayed an optical spectral transition from a \hbox{type 1} quasar to a type 1.9 AGN (no broad $\rm H\beta$ line) from 2000 to 2010. The corresponding Eddington ratio ($\lambda_{\rm Edd}$)\footnote{$\lambda_{\rm Edd}=L_{\rm bol}$/$L_{\rm Edd}$, where $L_{\rm bol}$ is the bolometric luminosity, $L_{\rm Edd} \approx 1.3 \times 10^{38}~M_{\rm BH}/\msun~\rm erg~s^{-1}$ is the Eddington luminosity, and $M_{\rm BH}$ is the SMBH mass.} dropped from $\sim 4\%$ to $\sim 0.7\%$. The serendipitous \xray\ observations  by {\it XMM-Newton} in 2000 and \chandra\ in 2005 revealed that the \hbox{2{--}10 keV} flux dropped by a factor of $\approx7.2$  with no clear evidence of  \xray\ obscuration \citep{LaMassa2015}.

In the scenario of change of obscuration, the \xray\ absorber has variable covering factor and/or column density, resulting in  strong \xray\ variability. Such changes are usually identified by characteristic changes of the \xray\ spectral shape along with the strong flux variations
%, which is an indication of \xray\ obscuration 
\citep[e.g.,][]{Turner2009,Yang2016,Li2019}. Previously, most of these events were found in type 2 AGNs \citep[e.g.,][]{Guainazzi2002,Risaliti2002,Puccetti2004,Risaliti2010,Rivers2011,Braito2013,Markowitz2014,Marinucci2016,Hickox2018,Jana2022}, where the variable absorber might be the gas ``clumps'' in the dusty torus  or the broad-line region (BLR). More recently, there have also been reports of type 1 AGNs and quasars with strong \xray\ obscuration variability \citep[e.g.,][]{Miniutti2009, Saez2012,Nanni2018,Liu2019,Ni2020,Timlin2020A,Timlin2020,Liu2021,Liu2022}.
There was no contemporaneous strong optical/UV continuum or emission-line variability, and these AGNs generally have high accretion rates \citep[e.g.,][]{Liu2019,Liu2021,Ni2020}.
One interpretation is that the \hbox{dust-free} absorber is a clumpy accretion-disk wind launched via radiation pressure
 \citep[e.g.,][]{Baskin2014,Matthews2016,Giustini2019}.

TDEs may cause strong \xray\ variability in both AGNs and inactive galaxies \citep[e.g.,][]{Komossa2015,Saxton2021,Gezari2021,Zabludoff2021}. 
The transient accretion process of a TDE could produce luminous \xray\ emission that decays roughly following the mass fallback rate of the stellar debris with a typical timescale of months-to-years \citep[e.g.,][]{Saxton2021}.
TDEs are generally found in inactive galaxies. %, and the typical TDE \xray\ spectra can be described by single black-body ($kT\approx 10\textrm{--}100~\rm ev$) or steep power law ($\Gamma\gtrsim4$; e.g., \citealt{Saxton2021}). 
There are only a few AGN TDE candidates proposed  \citep[e.g.,][]{Brandt1995,Merloni2015,Zhang2019,Liu2020,Ricci2020,Zhang2021}, as it is generally more difficult to identify TDEs in AGNs due to the luminous  persistent  AGN radiation \citep[e.g.,][]{Zabludoff2021}.

In this paper, we present \xray\ and multiwavelength analyses of an unusual source that showed extreme \xray\ variability (0.5--2 keV  flux variation factor $> 12.1$)  in the \hbox{7 Ms} \chandra\ Deep \hbox{Field-South} (CDF-S). It was identified as \xid\ in the \cite{Luo2017} \xray\ source catalog.   
This paper is organized as follows. The basic source properties and the \xray\ data are described in Section \ref{sec:data}. Section \ref{sec:results} presents the \xray\ light curve, \xray\ spectral analysis, optical/UV variability analysis, and multiwavelength properties. In  Section \ref{sec:discussion}, we discuss  possible explanations for the extreme \xray\ variability of XID 403. We summarize our results in Section \ref{sec:summary}. Throughout this paper,  we use a Galactic hydrogen column density of \hbox{$N_{\rm H} = 6.42 \times 10^{19}~\rm cm^{-2}$} \citep{HI4PI} toward the direction of \xid. Uncertainties and upper/lower limits  are quoted at 1$\sigma$ and 90\% confidence levels, respectively.
We adopt a $\Lambda\rm CDM$ cosmology with $\Omega_\Lambda = 0.685$, $\Omega_{\rm M} = 0.315$, and \hbox{$H_0 = 67.4~\kms~\rm Mpc^{-1}$} \citep{Planck2020}.

\section{Basic  object properties and X-ray Data}
\label{sec:data}
\subsection{Basic source properties and classification}\label{classification}\label{prop}

\begin{figure}
\centerline{
\includegraphics[width=3.5in]{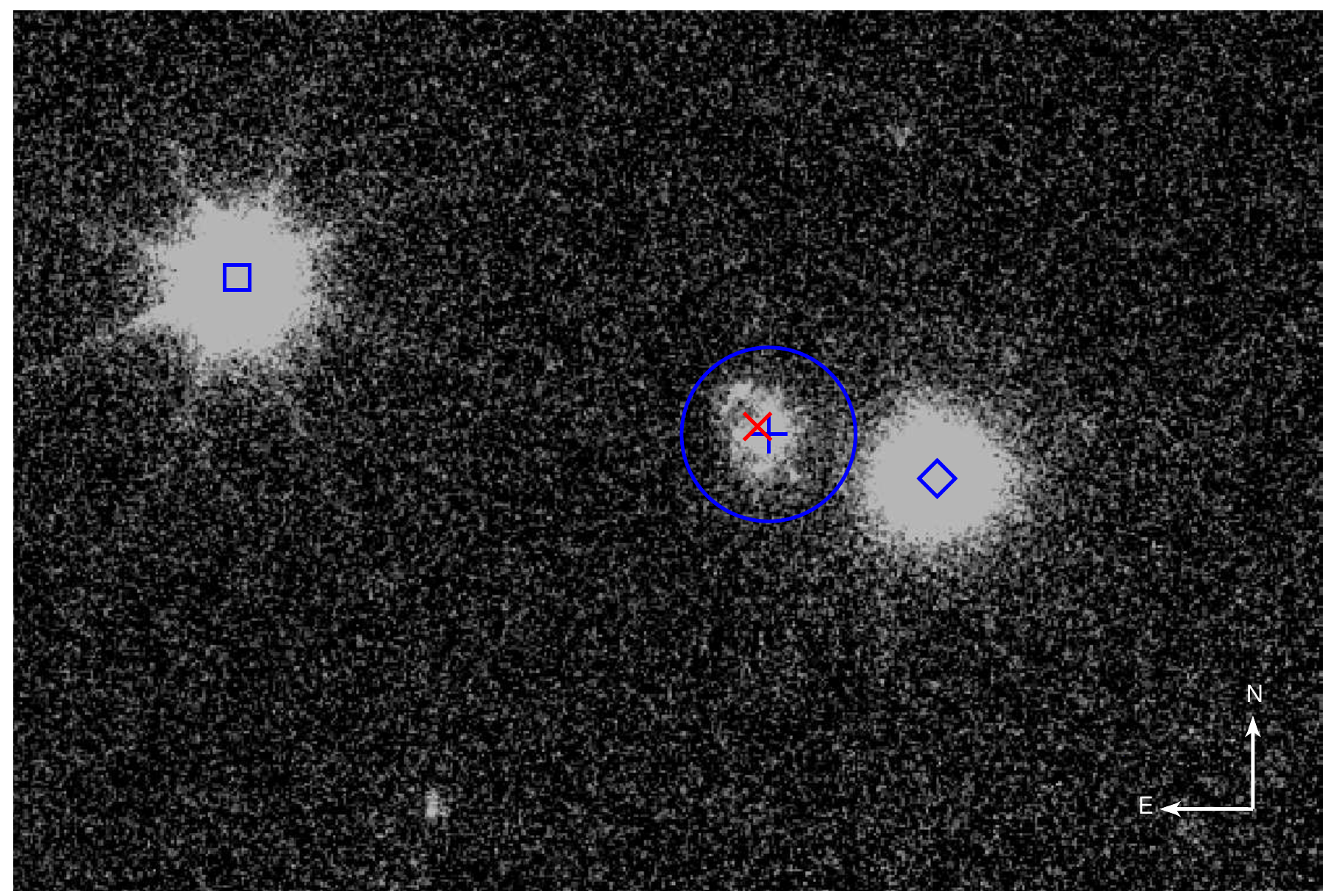}}
\caption{{\it HST} $z_{850}$-band image ($17\arcsec\times10\arcsec$) in the vicinity of \xid. 
The red ``x'' denotes the \xray\ position of \xid. The blue circle is centered on the optical position (the blue cross) with a radius of 1\arcsec, which is used for the aperture photometry in Section~\ref{sec2.2} below. The blue diamond and the blue square denote the nearby brighter galaxy and a nearby star (see Section \ref{sec2.2} below), respectively.}
\label{opim}
\end{figure}

\begin{figure*}
\centerline{
\includegraphics[width=5.0in]{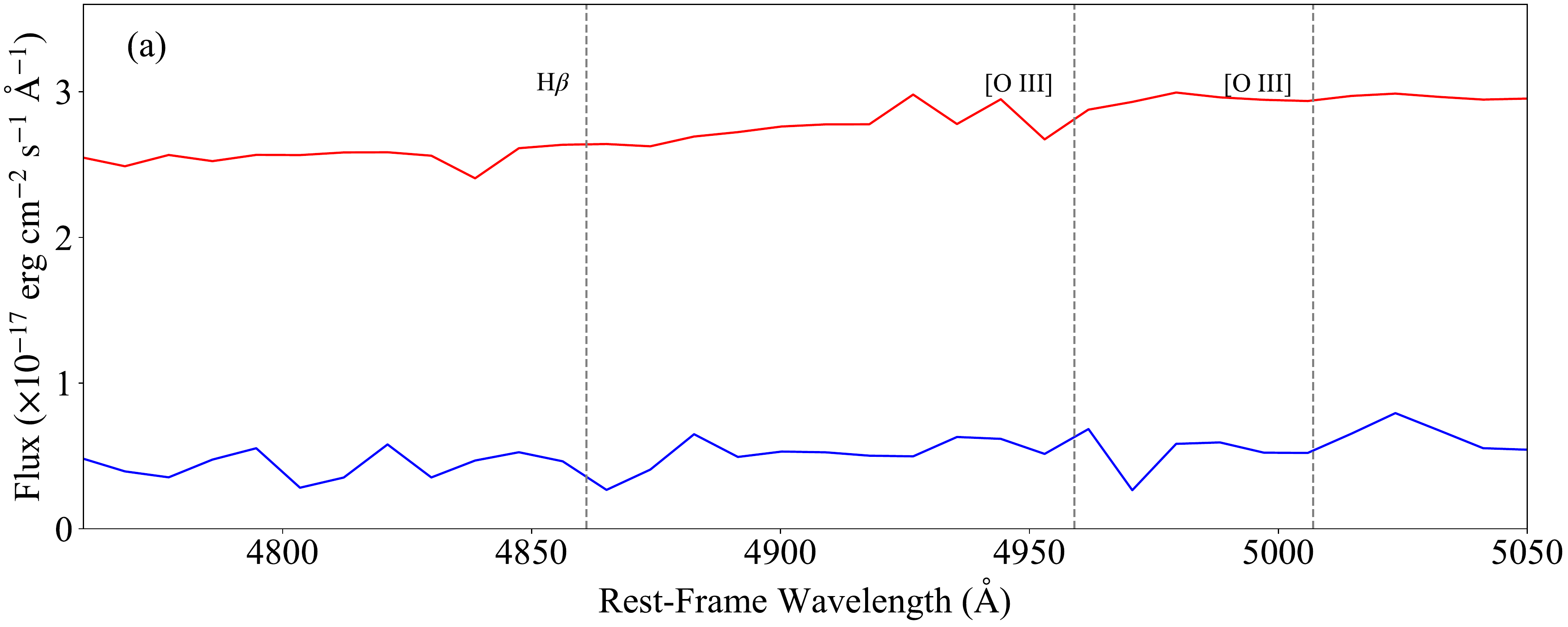}}
\centerline{
\includegraphics[width=5.0in]{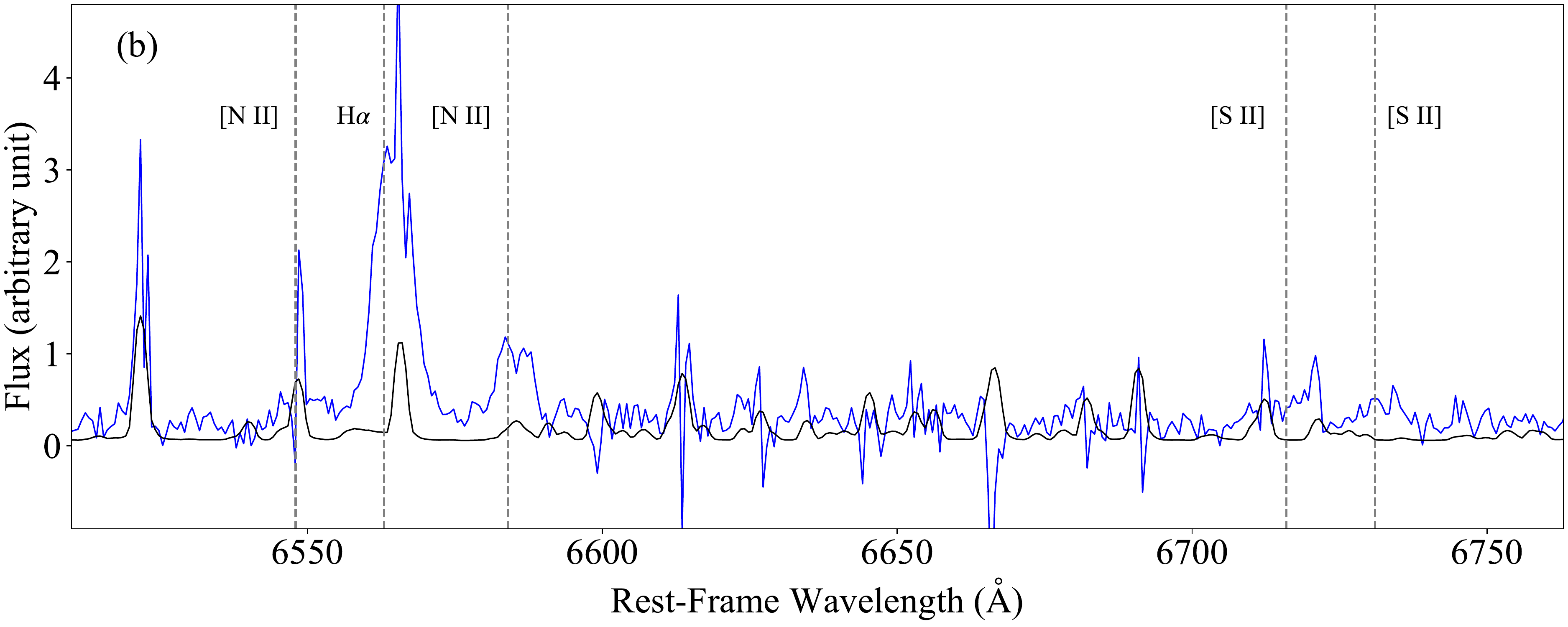}}
\caption{NIR spectra of \xid. Panel (a)  is the 3D-HST  spectrum covering the H$\beta$ line. 
The blue curve is the contamination-subtracted fluxes, and the red curve shows the contamination from the overlapping spectrum of the nearby galaxy. The spectrum uncertainties (at the $0.1\times10^{-17}\rm~erg~cm^{-2}~s^{-1}~\AA^{-1}$  level) are negligible compared to the contamination.
Panel (b) is the Keck/MOSFIRE spectrum covering the H$\alpha$ line. 
The blue and black curves are the spectrum and its errors, respectively. 
In both panels, the vertical dashed lines mark the locations of the main emission lines.  The emission feature left to the [\ion{N}{2}] $\lambda$6548 line has large uncertainties and is probably not real.
}
\label{opspec}
\end{figure*}

XID 403  was first reported by \cite{Luo2014} as a TDE candidate due to its \xray\ brightening and previous non-detection. 
Its basic \xray\ properties were then presented in the 7 Ms \hbox{CDF-S} source catalog \citep{Luo2017}. The J2000 \xray\ position is \hbox{$\rm RA = 53.094719$}, \hbox{$\rm DEC = -27.694609$} with a $1\sigma$ positional uncertainty of 0.38\arcsec. The off-axis angle is 6.80\arcmin, and the number of net source counts over the entire exposure is $\approx255$ in the \hbox{0.5{--}7 keV} band. The source appears soft, as it was not detected in the 2--7 keV band considering the entire exposure (net counts $<52.9$ at a 90\% confidence level).

The extreme \xray\ variability of \xid\ has also been noted in other previous studies. Its \xray\ light curve was presented in \cite{Zheng2017}, and it was considered a TDE candidate. In \cite{Paolillo2017}, it was reported to have an extremely large excess variance ($\approx 3$) 
compared to other CDF-S sources with similar \xray\ counts.

The optical counterpart of \xid\ is a galaxy (centroid \hbox{$\rm RA = 53.094679$}, \hbox{$\rm DEC = -27.694634$}) that is 0.16\arcsec~away from the \xray\ position.  A Hubble Space Telescope ({\it HST}) $z_{850}$-band image is shown in Figure \ref{opim}, and the host galaxy is slightly extended.
Considering the \xray\ position uncertainty of 0.38\arcsec, \xid\ is consistent with being located at the center of the host galaxy. The host has AB magnitudes of \hbox{$R = 24.4$}, \hbox{$z_{\rm 850} = 23.5$}, and \hbox{$K_s = 21.5$}. To the southwest, there is a brighter galaxy at $z = 0.535$ with a separation of only 1.9\arcsec; its \rband\ flux is 5.7 times higher than that of the \xid\ host \citep{Straatman2016}. This nearby galaxy is not an \xray\ source, but it may affect the optical or near-infrared (NIR) measurements of the host properties. 

From a Keck/MOSFIRE NIR spectrum of the host galaxy, which shows clear H$\alpha$, [\ion{N}{2}], and [\ion{S}{2}]  narrow emission lines, \cite{Trump2013} identified a spectroscopic redshift of $z=1.608$. Combining this spectrum and a NIR spectrum from the {\it HST} Wide Field Camera 3 (WFC3) G141 slitless grism observation obtained as a part of the \hbox{3D-HST} survey \citep{Brammer2012,Momcheva2015}, which covers the H$\beta$ and [\ion{O}{3}]  lines,  \cite{Trump2013} also proposed that the object is a Seyfert 2 galaxy based on the classic ``BPT'' and ``VO87''  emission-line diagnostics \citep{BPT,VO87}. The 3D-HST spectrum \citep{Momcheva2015} and the Keck/MOSFIRE spectrum  \citep{Trump2013}   are plotted in Figure \ref{opspec}. The 3D-HST spectrum has a low spectral resolution ($\lambda/\Delta \lambda\approx130$), and it is severely contaminated by the overlapping spectrum of the nearby galaxy. Therefore, the H$\beta$ and [\ion{O}{3}] measurements are highly 
uncertain. Thus, the Seyfert 2 classification 
proposed by \cite{Trump2013} is  unreliable.

We examine the Keck/MOSFIRE spectrum for AGN 
signatures. 
We fit this spectrum using the the Python package {\sc PyQSOFit} \citep[e.g.,][]{Guo2019,Shen2019,Wang2019}. The best-fit full width at half-maximum (FWHM) of the H$\alpha$ line profile is \hbox{$\approx 350~\rm km~s^{-1}$}, significantly smaller than the broad-line criteria (e.g., \hbox{$\gtrsim800~\rm km~s^{-1}$}). 
The best-fit $\log$[\ion{N}{2}]/H$\alpha$ value is $-0.54$, consistent with the measurement ($-0.6$) 
in \cite{Trump2013}. 
The rest-frame equivalent width of H$\alpha$
is $W_{\rm H\alpha} \approx 70$~\AA. These do not
satisfy the AGN criterion based on the 
$W_{\rm H\alpha}$ versus [\ion{N}{2}]/H$\alpha$ (WHAN) diagnostics ($\log$[\ion{N}{2}]/H$\alpha>-0.4$ and $W_{\rm H\alpha}>3$~\AA; 
e.g., \citealt{CidFernandes2011}). Therefore, there is no clear AGN signature in the available NIR spectra of this \hbox{high-redshift} faint galaxy.
%and the Seyfert 2 classification are highly unreliable, there is thus no clear AGN signature in the available spectra of this \hbox{high-redshift} faint galaxy.

Nevertheless, \xid\ appears to be a luminous \xray\ source. At $z=1.608$, its rest-frame 0.5--7 keV  luminosity reaches \hbox{$\approx 2.1 \times 10^{43}~\rm erg~s^{-1}$}  over the entire 7 Ms exposure (\citealt{Luo2017}; updated to the spectroscopic redshift). Such a large amount of power should originate from SMBH accretion. Thus \xid\ is either an AGN or associated with TDEs.

\subsection{X-ray Data}\label{xdata}
 With a co-added \chandra\ depth of $\approx$ 7 Ms, the CDF-S is the deepest \xray\ survey to date  \citep{Luo2017,Xue2017}. The \xray\ data consist of 102 observations collected by the \chandra\ Advanced CCD Imaging Spectrometer imaging array \citep[\hbox{ACIS-I;}][]{Garmire2003}. We use the cleaned event files from \cite{Luo2017} for data analyses. 
 For a given observation or observational epoch (described below), we use ACIS Extract (AE; \citealt{Broos2010}) to extract the \xray\ source counts in the full band (\hbox{0.5--5 keV}), soft band (\hbox{0.5--2 keV}), and hard band (\hbox{2--5 keV}). We adopt an upper energy bound of \hbox{5 keV} because the \xray\ spectrum of \xid\ is very soft and the $> 5$ keV spectrum is dominated by background.
 In the source rest frame, the full band corresponds to the \hbox{1.3--13 keV} band, probing the hard \hbox{X-rays}. We use the binomial \hbox{no-source} probability (\pb; e.g., \citealt{Broos2007,Xue2011,Luo2013,Luo2015,Xue2016}) calculated by AE to determine the significance of the source signal in each band. We adopt \hbox{\pb = 0.01} as the detection threshold; this is appropriate for a source with a pre-specified position.  If $\pb < 0.01$, we consider the source detected and provide measurements of the source counts. The 1$\sigma$ uncertainties of the counts are computed following the  \cite{Gehrels1986} approach. If $\pb \ge 0.01$, we consider the source undetected and provide a 90\% confidence-level  upper limit on the source counts using the Bayesian approach described in \cite{Kraft1991}.
 
XID 403 was significantly detected in the latest 3 Ms exposure of the 7 Ms \hbox{CDF-S}.
To investigate if the source was detected in any of the previous observations,
we examine the \pb values in each observation of the first 4 Ms exposure. We find one  observation  (observation ID 8595) in the \hbox{second Ms} exposure, in which \xid\ was significantly detected in both the full and soft bands with corresponding \pb values of 0.001 and 0.0004, respectively. The first 4 Ms exposure consists of 54 individual observations, and the expected number of false detections with $\pb \le 0.0004$ is estimated to 
be $\lesssim0.02 (54 \times 0.0004)$. 
Therefore, we consider the source marginally detected (at a $\approx2.1\sigma$ significance level) in this observation.  We also check the observations adjacent to observation ID~8595. Although they do not provide individual detections, we find that combining the two observations prior to observation ID~8595 (observations IDs 8593 and 8597) leads to a more robust detection. \xid\ was significantly detected in all three bands in the combined observation, with $\pb$ values of $2.6\times10^{-5}$, $4.3\times10^{-4}$, and 0.009 in the full, soft, and hard bands, respectively.
The expected number of false detections with $\pb\le2.6\times10^{-5}$ from such a combined observation (combined from three consecutive observations) is estimated to be $\lesssim 0.0013$ [$(54 - 2)\times2.6\times10^{-5}$], corresponding to a $\approx3.0\sigma$ significance level.\footnote{We also investigate via simulations whether the significance enhancement (from $2.1\sigma$ to $3.0\sigma$) is due to background fluctuations in the two additional observations. We assume that these two observations contain only background counts, and we simulate the extracted counts by drawing randomly from their respective Poisson distributions (similar to the method described in Section~\ref{sec:LC} below). The resulting fraction of cases where the full-band $\pb$ value in the combined observations reaches $<2.6\times10^{-5}$ is only  0.0049. This indicates that the significance enhancement is unlikely due to background fluctuations. The probability of \xid\ not being detected in either observation ID~8595 or the combined observation is thus $0.02 \times0.0049 + 0.0013 \approx 0.0014$ (the former term for the detection in observation ID~8595 being spurious plus the significance enhancement from the combined observation being also spurious and the latter term for the detection in the combined observation being spurious), still corresponding to  $\approx 3.0\sigma$ detection significance.   }
This combined observation is thus used in our following analyses.

In order to investigate \xray\ variability, we group the \xray\ data into 6 epochs according to the \xray\ flux states and  observational dates.
The 11 observations in the first Ms exposure are combined as \hbox{epoch 1}. In the second Ms exposure, the three observations before observation ID 8593 are combined  as epoch 2, observation  IDs 8593, 8597, and 8595 are combined as epoch 3, and the six observations after observation ID 8595 are combined as epoch 4. We note that the second Ms exposure was performed within $\approx1.5$ months, and thus epoch 2--4 are close in time, allowing us to investigate the rapid variability  for epoch~3 (see Section \ref{sec:LC} below). 
%and the six observations after observation ID 8595 are combined as \hbox{epochs 2} and 4, respectively. 
The 31 observations in the third and fourth Ms exposures are combined as epoch 5. The 48 observations in the latest 3 Ms of exposure are combined as \hbox{epoch 6}. In epochs 1, 2, 4, and 5, XID 403 was not detected in any of the three bands in either the  individual observations or the combined observations. 
In epochs 3 and 6, the source was detected in  all three bands. The basic information and  \xray\ photometric properties  for the six epochs    are listed in Table \ref{tab1}.
To investigate \hbox{shorter-term} \xray\ variability in epoch 6, we further divide epoch 6 into six subepochs with comparable exposure times. The subepoch information is shown in \hbox{Table \ref{tab2}}.

To assess the spectral shapes,
we derive $\Gamma_{\rm eff}$ for epoch 3, epoch 6, and the subepochs of epoch 6. Assuming a simple power-law spectrum that is modified by the Galactic  absorption,
we derive $\Gamma_{\rm eff}$ values from the band ratios, defined as the ratio between the hard-band and soft-band net counts,
following the approach described in Section 4.4 of \cite{Luo2017}.  XSPEC (v12.11.1; \citealt{Arnaud1996}) and the spectral response files are used in this procedure.  For the epochs/subepochs where \xid\ is  detected in both the soft and hard bands,  the  $\Gamma_{\rm eff}$ uncertainties were propagated from the uncertainties of the net counts. If \xid\ is detected in only the soft band but not the hard band, we derive a lower limit on   $\Gamma_{\rm eff}$  from the upper limit on the band ratio.
The $\Gamma_{\rm eff}$ constraints are listed in Tables \ref{tab1} and \ref{tab2}.

XID 403 has been covered by other \chandra\ and \hbox{\it XMM-Newton} surveys. We match the position of \xid\ to the $\approx 3$ Ms depth XMM deep survey in the \hbox{CDF-S} (\hbox{XMM-CDFS}) source catalog \citep{Ranalli2013}, the 250 ks depth \chandra\ extended \hbox{CDF-S} (\hbox{E-CDF-S}) catalog \citep{Lehmer2005,Xue2016}, and the $\approx 30$ ks depth XMM-Spitzer Extragalactic Representative Volume Survey (\hbox{XMM-SERVS}) catalog \citep{Ni2021}, and \xid\ was not detected in any of these surveys.
The XMM-CDFS consists of \hbox{33 observations}. The first \hbox{8 observations} have a total exposure of $\approx 0.5$ Ms, and they were performed between epochs 1 and 2. The latter 25 observations with a total  exposure of $\approx 2.5$ Ms were performed between epochs 4 and 5. 
The flux upper-limit constraints from the $\approx 0.5$ Ms and $\approx 2.5$ Ms {\it XMM-Newton} exposures are not as stringent as those of  epoch 1 and epoch 5, which have \hbox{$\approx 1$ Ms} and $\approx 2$ Ms \chandra\ exposures, respectively. 
The E-CDF-S observations were performed between epochs 1 and 2,  and the  effective exposure time is lower than those of epochs 1 and 2. Thus, the non-detection is not constraining either.
\xid\ probably remained in a low state during the E-CDF-S and XMM-CDFS observations, and thus it was not detected. 
The \hbox{XMM-SERVS} observations were carried out after \hbox{epoch 6}. The non-detection of \xid\ indicates a \hbox{0.5--10 keV} flux upper limit of \hbox{$\approx 1.0 \times 10^{-14}~\rm erg~cm^{-2}~s^{-1}$}, which is \hbox{$\approx 6.4$ times} larger than the epoch 6 flux (derived from the epoch 6 spectral fitting; see Section \ref{sec3.2} below). Thus the XMM-SERVS survey does not provide useful constraints either due to its shallower depth.
We do not use these \xray\ data in the following analyses.

\section{X-ray and Multiwavelength Properties} \label{sec:results}
\subsection {X-ray Light Curve}\label{sec:LC}

\begin{figure}
\includegraphics[width=3.3in]{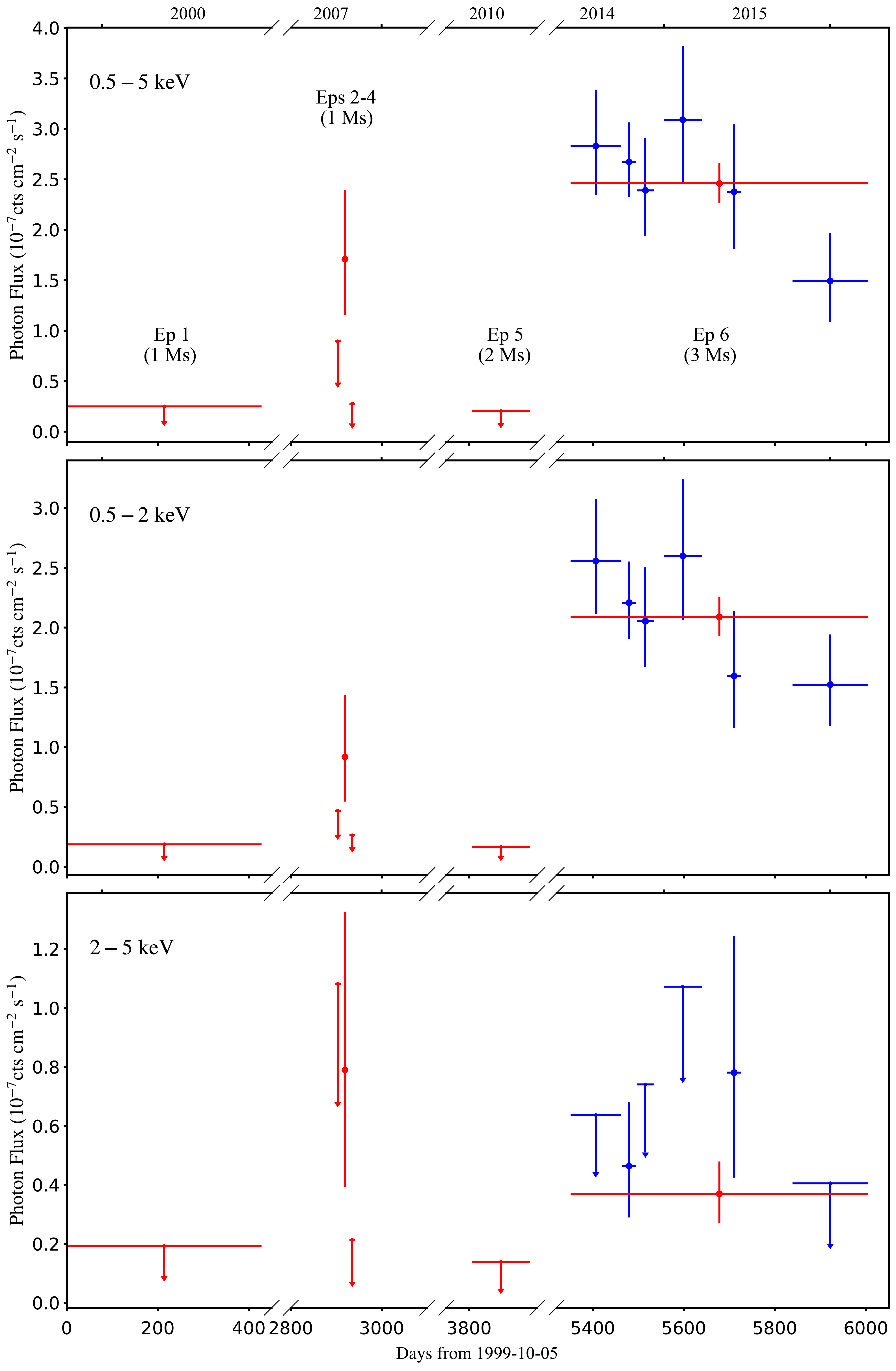}
\caption{\xray\ light curves of XID 403 in the full band (top panel), soft band (middle panel), and hard band (bottom panel). The PFs of the six epochs are shown in red, and the PFs of the six subepochs in epoch~6 are shown in blue.
%Epoch 6 is divided into six subepochs.
The data points with vertical error bars indicate the detections with $1\sigma$ uncertainties, and the arrows indicate upper limits. 
The horizontal error bars indicate the bin sizes of the time bins.}
\label{xlc}
\end{figure}

\tabletypesize{\scriptsize}
\tablewidth{0pt}
\begin{deluxetable*}{crrrrrrrrrrc}
\tablecaption{X-ray Photometric Properties in the Six Observational Epochs}
\label{tab1}
\tablehead{
\colhead{}&
\colhead{}&
\colhead{}&
\colhead{}&
\colhead{}& 
\multicolumn{2}{c}{Full Band (0.5{--}5 keV)}   & 
\multicolumn{2}{c}{Soft Band (0.5--2 keV) }   & 
\multicolumn{2}{c}{Hard Band (2--5 keV) }&
\colhead{}\\ 
\cline{6-11}
\colhead{Epoch}&
\colhead{Obs. ID} &
\colhead{Obs. Date} & 
\colhead{Time} &
\colhead{Exposure} &
\colhead{ Net} &   
\colhead{Photon} & 
\colhead{Net} & 
\colhead{Photon} & 
\colhead{Net} & 
\colhead{Photon}   & 
\colhead{$\Gamma_{\rm eff}$}     \\
\colhead{Number}   & 
\colhead{Range} &  
\colhead{Range}       & 
\colhead{Interval}    &
\colhead{Time (ks)}        &
\colhead{Counts}    & 
\colhead{Flux}&    
\colhead{Counts} & 
\colhead{Flux} &
\colhead{Counts}  &
\colhead{Flux}& 
\colhead{}\\
\colhead{(1)}&
\colhead{(2)}&
\colhead{(3)}&
\colhead{(4)}&
\colhead{(5)}&
\colhead{(6)}&
\colhead{(7)}&
\colhead{(8)}&
\colhead{(9)}&
\colhead{(10)}&
\colhead{(11)}&
\colhead{(12)}
}
\startdata
1& 441\textrm{--}2409 & 1999-10-15\textrm{--}2000-12-16 & 0\textrm{--}428 & 931 &$ <10.1$&$ <0.25$&$ <7.6$&$ <0.19$&$ <7.7$&$ <0.19$&--\\
2 & 8591\textrm{--}9718 & 2007-09-20\textrm{--}2007-10-03 & 2897\textrm{--}2910 & 138 &$ <4.4$&$ <0.90$&$ <2.3$&$ <0.47$&$ <5.3$&$ <1.08$&--\\
3 & 8593\textrm{--}8597 & 2007-10-06\textrm{--}2007-10-19 & 2913\textrm{--}2926 & 222 &$ 14.3^{+5.8}_{-4.6}$&$ 1.71^{+0.68}_{-0.55}$&$ 7.7^{+4.3}_{-3.1}$&$ 0.92^{+0.52}_{-0.38}$&$ 6.6^{+4.5}_{-3.3}$&$0.79^{+0.54}_{-0.40}$&$1.4_{-0.7}^{+0.9}$\\
4& 8592\textrm{--}9596 & 2007-10-22\textrm{--}2007-11-04 & 2929\textrm{--}2942 & 600 &$ <6.4$&$  <0.28$&$ <6.1$&$ <0.26$&$ <4.9$&$ <0.21$&--\\
5& 12043\textrm{--}12234 & 2010-03-18\textrm{--}2010-07-22 & 3807\textrm{--}3933 & 1956 &$ <15.7$&$ <0.20$&$ <12.8$&$ <0.17$&$ <10.7$&$ <0.14$&--\\
 6&16176\textrm{--}18730&2014-06-09\textrm{--}2016-03-24 & 5351\textrm{--}6005 & 2881 &$ 231.6^{+18.6}_{-17.6}$&$ 2.46^{+0.20}_{-0.19}$&$ 197.0^{+16.2}_{-15.1}$&$ 2.09^{+0.17}_{-0.16}$&$ 34.6^{+10.0}_{-8.9}$&$ 0.37^{+0.11}_{-0.10}$&$3.0^{+0.4}_{-0.3}$\\
\enddata
\tablecomments{Column (1): epoch number. Column (2): range of the \chandra\ observation IDs; the observations were not carried out following strictly the order of the observation IDs, and thus a later epoch might contain smaller observation IDs. Column (3): range of the observation start date. Column (4): time interval in units of days, starting from the beginning of the CDF-S observations. Column (5): total exposure time in units of ks. Columns (6), (8), and (10): full-, soft-, and hard-band net source counts;  for non-detections, 90\% confidence-level upper limits are given. Columns (7), (9), and (11): full-, soft-, and hard-band photon fluxes in units of $\rm 10^{-7}~cts~cm^{-2}~s^{-1}$;  for non-detections, 90\% confidence-level upper limits are given.
}

\end{deluxetable*}

\tabletypesize{\scriptsize}
\tablewidth{0pt}
\begin{deluxetable*}{crrrrrrrrrrr}
\label{tab2}
\tablecaption{Subepoch X-ray Photometric Properties in Epoch 6}
\tablehead{
\colhead{}&
\colhead{}&
\colhead{}&
\colhead{}&
\colhead{}& 
\multicolumn{2}{c}{Full Band (0.5{--}5 keV)}   & 
\multicolumn{2}{c}{Soft Band (0.5--2 keV) }   & 
\multicolumn{2}{c}{Hard Band (2--5 keV) }&
\colhead{}\\ 
\cline{6-11}
\colhead{Subepoch}&
\colhead{Obs. ID} &
\colhead{Obs. Date} & 
\colhead{Time} &
\colhead{Exposure} &
\colhead{Net} &   
\colhead{Photon} & 
\colhead{Net} & 
\colhead{Photon} & 
\colhead{Net} & 
\colhead{Photon}   & 
\colhead{$\Gamma_{\rm eff}$}\\
\colhead{Number}   & 
\colhead{Range} &  
\colhead{Range}       & 
\colhead{Interval}    &
\colhead{Time (ks)}        &
\colhead{Counts}    & 
\colhead{Flux}&    
\colhead{Counts} & 
\colhead{Flux} &
\colhead{Counts}  &
\colhead{Flux}& 
\colhead{}\\
\colhead{(1)}&
\colhead{(2)}&
\colhead{(3)}&
\colhead{(4)}&
\colhead{(5)}&
\colhead{(6)}&
\colhead{(7)}&
\colhead{(8)}&
\colhead{(9)}&
\colhead{(10)}&
\colhead{(11)}&
\colhead{(12)}
}
\startdata
 1& 16180\textrm{--}17417 & 2014-06-09\textrm{--}2014-09-28 & 5351\textrm{--}5462 & 445 &$ 41.1^{+8.1}_{-7.0}$&$ 2.83^{+0.56}_{-0.48}$&$ 37.2^{+7.5}_{-6.4}$&$ 2.56^{+0.52}_{-0.44}$&$ <9.3$&$ <0.64$&$> 2.5$\\
 2& 16175\textrm{--}17542 & 2014-10-01\textrm{--}2014-10-31 & 5465\textrm{--}5495 & 826 &$ 71.0^{+10.4}_{-9.3}$&$ 2.67^{+0.39}_{-0.35}$&$ 58.6^{+9.1}_{-8.1}$&$ 2.21^{+0.34}_{-0.30}$&$ 12.3^{+5.7}_{-4.6}$&$ 0.46^{+0.22}_{-0.17}$&$2.8^{+0.6}_{-0.5}$\\
 3& 16186\textrm{--}17556 & 2014-11-02\textrm{--}2014-12-09 & 5497\textrm{--}5534 & 513 &$ 39.2^{+8.5}_{-7.4}$&$ 2.39^{+0.52}_{-0.45}$&$ 33.7^{+7.4}_{-6.3}$&$ 2.05^{+0.45}_{-0.39}$&$ <12.2$&$ <0.74$&$> 2.5$\\
 4& 16179\textrm{--}17573 & 2014-12-31\textrm{--}2015-03-24 & 5556\textrm{--}5639 & 295 &$ 31.7^{+7.4}_{-6.4}$&$ 3.09^{+0.73}_{-0.62}$&$ 26.6^{+6.6}_{-5.5}$&$ 2.60^{+0.64}_{-0.53}$&$ <11.0$&$ <1.07$&$> 2.0$\\
 5& 16191\textrm{--}16461 & 2015-05-19\textrm{--}2015-06-20 & 5695\textrm{--}5727 & 304 &$ 25.1^{+7.1}_{-6.0}$&$ 2.38^{+0.67}_{-0.56}$&$ 16.8^{+5.7}_{-4.6}$&$ 1.60^{+0.54}_{-0.43}$&$ 8.2^{+4.9}_{-3.8}$&$ 0.78^{+0.46}_{-0.36}$&$1.8^{+1.0}_{-0.6}$\\
 6& 16185\textrm{--}18730 & 2015-10-10\textrm{--}2016-03-24 & 5839\textrm{--}6005 & 496 &$ 23.5^{+7.5}_{-6.4}$&$ 1.49^{+0.48}_{-0.41}$&$ 24.0^{+6.6}_{-5.5}$&$ 1.52^{+0.42}_{-0.35}$&$ <6.4$&$ <0.41$&$> 2.6$\\
    \enddata
\tablecomments{The same as Table \ref{tab1}, but for the six subepochs in epoch 6.}

\end{deluxetable*}

%\tabletypesize{\scriptsize}

To construct \xray\ light curves of XID 403,
we calculate the \xray\ photon flux (PF) in each band and each epoch/subepoch as follows \citep[e.g.,][]{Yang2016,Ding2018}:
\begin{equation}
\label{eqpf1}
{\rm PF} = \frac{NET\_CNTS}{EFFAREA \times EXPOSURE \times PSF\_FRAC}.
\end{equation}
The $1\sigma$ uncertainty of PF is computed as:
\begin{equation}
{\rm \delta PF} = \frac{\delta NET\_CNTS}{EFFAREA \times EXPOSURE \times PSF\_FRAC}.
\label{eqpf2}
\end{equation}
In the above formulas, $NET\_CNTS$ and $\delta NET\_CNTS$ are the numbers of background-subtracted counts (i.e., net counts) and its uncertainty (Tables \ref{tab1} and \ref{tab2}).  
$EFFAREA$ is the  effective area calculated by AE, and it is the product of the mirror geometric area, reflectivity, off-axis vignetting, and detector quantum efficiency.
$EXPOSURE$ is the exposure time. $PSF\_FRAC$ is the PSF fraction of the source region ($\approx89\%\textrm{--}91\%$). For the non-detections, the corresponding PF upper limits are computed.
The PF constraints are listed in Tables \ref{tab1} and \ref{tab2}. 

Figure \ref{xlc} displays the full-, soft- and hard-band light curves.
In the full and soft bands, \xid\ was only detected in epoch 3, epoch 6, and all the subepochs of epoch 6.  
%\xid\ showed two \xray\ brightening events.
In \hbox{epochs 1} and 2, it was in a low state. 
In  epoch 3, the source brightened to an intermediate state; the separation between the median dates of  epochs 2 and 3  is 6.1 days in the rest frame. \xid\ then returned to a low state in \hbox{epoch 4}; the separation between epochs 3 and 4 is also 6.1 rest-frame days. 
%The rise and decay \textbf{time} for the \hbox{epoch 3} outburst are both short. The duration of the outburst appears also short, which has a 
The duration of the outburst is
 $\approx5.0$ to 7.3 rest-frame days, 
where the lower bound is derived from the epoch~3 exposure time and the 
upper bound is computed from the end time of epoch 2 to the start
time of epoch 4. We note that the duration is constrained directly from the epoch~2--4 light curve (Figure~\ref{xlc}), and it does not necessarily represent the actual timescale of the outburst.\footnote{For example, epoch 2 could potentially belong to the rise phase of a longer-duration outburst but the \xray\ non-detection hampers its identification.}
In \hbox{epoch 5}, \xid\ was still in a low state.
In epoch 6, \xid\ brightened to a high state; the start time of the brightening is uncertain, which could be as early as the end time of epoch 5.  The source remains bright until the end of epoch 6; thus, the epoch 6 outburst has a duration of $>251$ rest-frame days.
We also estimate the time separation between the epoch 3 and \hbox{epoch 6} outbursts,
which has a range of $\approx1.1$ to 2.5 rest-frame years, where the lower bound
is computed from the start time of epoch 4 to the end time of epoch 5 and the upper bound is from the 
end time of epoch 3 to the start
time of epoch 6. Considering the long observational gaps between epochs 4 and 5 and between epochs 5 and 6, there could have been additional outbursts missed by the 7~Ms CDF-S, and thus the separation between the epoch~3 and epoch~6 outbursts does not necessarily  reflect the separation of outbursts in general.
The outburst properties are summarized in Table~\ref{tabprop}.

To compare the PF measurements to the upper limits in different epochs, we adopt
a Monte Carlo approach. 
For each epoch, we generate the probability density function (PDF) for the 
net source counts via Monte Carlo simulations of 
Poisson distributions for the extracted source and background counts.
The counts PDFs are then converted to the PF PDFs, and the 90\% confidence-level
lower limit on the flux variation factor between two epochs is then 
determined from the PDF of the ratio of the two PFs. 
Compared to epoch 2, the epoch 3 full- and soft-bands PF increased by factors of $>2.5$ and $>2.0$, respectively. In the hard-band, the epoch 3 PF is not larger than the epoch 2 PF at the
90\% confidence level.
Compared to epoch 3, the epoch 4 full-, soft-, and hard-bands PF
decreased by  factors of $>6.0$,  $>3.3$, and $>3.5$, respectively.
Compared to epoch 5, the \hbox{epoch 6}  PF increased by factors of \hbox{$> 12.6$}, \hbox{$> 12.1$}, \hbox{$> 3.1$}  in  the full,  soft, and hard bands, respectively. The brightening is more significant if considering subepoch 1 compared to epoch 5, with the PF  increasing by a factor of $> 13.9$ in the  full band ($> 15.1$ in the soft band). The full-band variability factors are summarized in Table~\ref{tabprop}. Such large \xray\ flux variability factors are rare among typical AGN populations \citep[e.g.,][]{Yang2016,Timlin2020}, making XID 403 an exceptional object.

\tabletypesize{\normalsize}
\tablewidth{0pt}
\begin{deluxetable*}{cccccc}
\label{tabprop}
\tablecaption{Outburst Properties}
\tablehead{
\colhead{}&
\colhead{Rise Time}&
\colhead{Duration}&
\colhead{Decay Time}&
\colhead{Flux State}& 
\colhead{Var. Factor \tablenotemark{a}}
}
\startdata
 Epoch 3& $\lesssim 6.1$~days & $\approx5.0$--7.3~days\tablenotemark{b} &$\lesssim6.1$ days&Intermediate&$>6.0$\\ 
 \midrule
%\multicolumn{6}{c}{\multirow{2}{*}{Separation $\approx1.1\textrm{--}2.5~\rm yrs$}} \\
%\multicolumn{6}{l}{} \\
\multicolumn{6}{c}{Separation: $\approx1.1\textrm{--}2.5$~years\tablenotemark{c}} \\
 \midrule
  Epoch 6& \multicolumn{1}{c}{--} & $>251$ days&\multicolumn{1}{c}{--}&High&$>12.6$\\ 
    \enddata

\tablenotetext{a}{
Full-band PF variability factor. For the epoch~3 outburst, it is for the decay phase.}
\tablenotetext{b}{The duration of the epoch 3 outburst is constrained directly from the epoch 2--4 light curve (Figure \ref{xlc}), and it does not necessarily represent the actual timescale of the outburst (e.g., see Footnote 4).}
\tablenotetext{c}{
Separation between epochs 3 and 6, there could have been additional outbursts missed by the 7 Ms CDF-S.}

\end{deluxetable*}

We examine if there is significant flux variability within epoch 6.
We utilize the $\chi^2$ approach (e.g., \citealt{Young2012,Yang2016,Paolillo2017,Ding2018}) to quantify the \xray\ variability. The $\chi^2$ values of the epoch 6 light curves are calculated as follows :
\begin{equation}
\chi^2 = \sum_{ i =1}^{6} \frac{({\rm PF}_{i} - \langle {\rm PF} \rangle)^2}{(\delta {\rm PF}_{i})^2},
\label{eqx2}
\end{equation}
where $\langle {\rm PF} \rangle$ is the  combined epoch 6 PF.  The \hbox{full-band} $\chi^2$ value is 6.6 and the soft-band value is 5.1.
To determine the significance of any PF variability, we use Monte Carlo simulations to estimate the probability ($P_{\chi^2}$) that the computed
 $\chi^2$ value is generated from the PF distribution of a constant intrinsic PF ($\langle {\rm PF} \rangle$) modified by Poisson noise \citep[e.g.,][]{Paolillo2004,Ding2018}. 
 Given the $\langle {\rm PF} \rangle$ value,  we compute the model source and background counts in each subepoch. Then we simulate the observed source and background counts by drawing randomly from their respective Poisson distributions  (see  detailed description in Section 3.1 of \citealt{Ding2018}).  We derive the simulated net counts and their uncertainties following the same procedure described in \hbox{Section \ref{xdata}}, and calculate the corresponding PF and $\chi^2$ values. 
 We perform 10\,000 simulations and obtain 10\,000 simulated $\chi^2$ values. The probability $P_{\chi^2}$ is computed as the fraction of  the simulations where the simulated $\chi^2$ value is larger than the observed value.  We obtain the $P_{\chi^2}$ values of  0.21 ($\approx 1.2 \sigma$) and 0.33 ($\approx 1.0 \sigma$) for the full and soft bands, respectively. These indicate that the observed $\chi^2$ values are likely due to Poisson fluctuations, and  there is no significant flux variability within epoch 6. 
 
 In epoch 6, the spectral shape is very soft, with a $\Gamma_{\rm eff}$ value of $3.0_{-0.3}^{+0.4}$.
 We investigate if there is variability of the \xray\ spectral shape within epoch 6. We compare the $\Gamma_{\rm eff}$ constraints in the six subepochs. The two $\Gamma_{\rm eff}$ measurements in subepochs 2 and 5 are consistent with each other within the uncertainties, and they also agree with the upper-limit constraints in the other subepochs.
 Thus there is no apparent spectral shape evolution within epoch 6.

\subsection {X-ray Spectral Analysis}\label{sec3.2}

\begin{figure}
\includegraphics[width=3.3in]{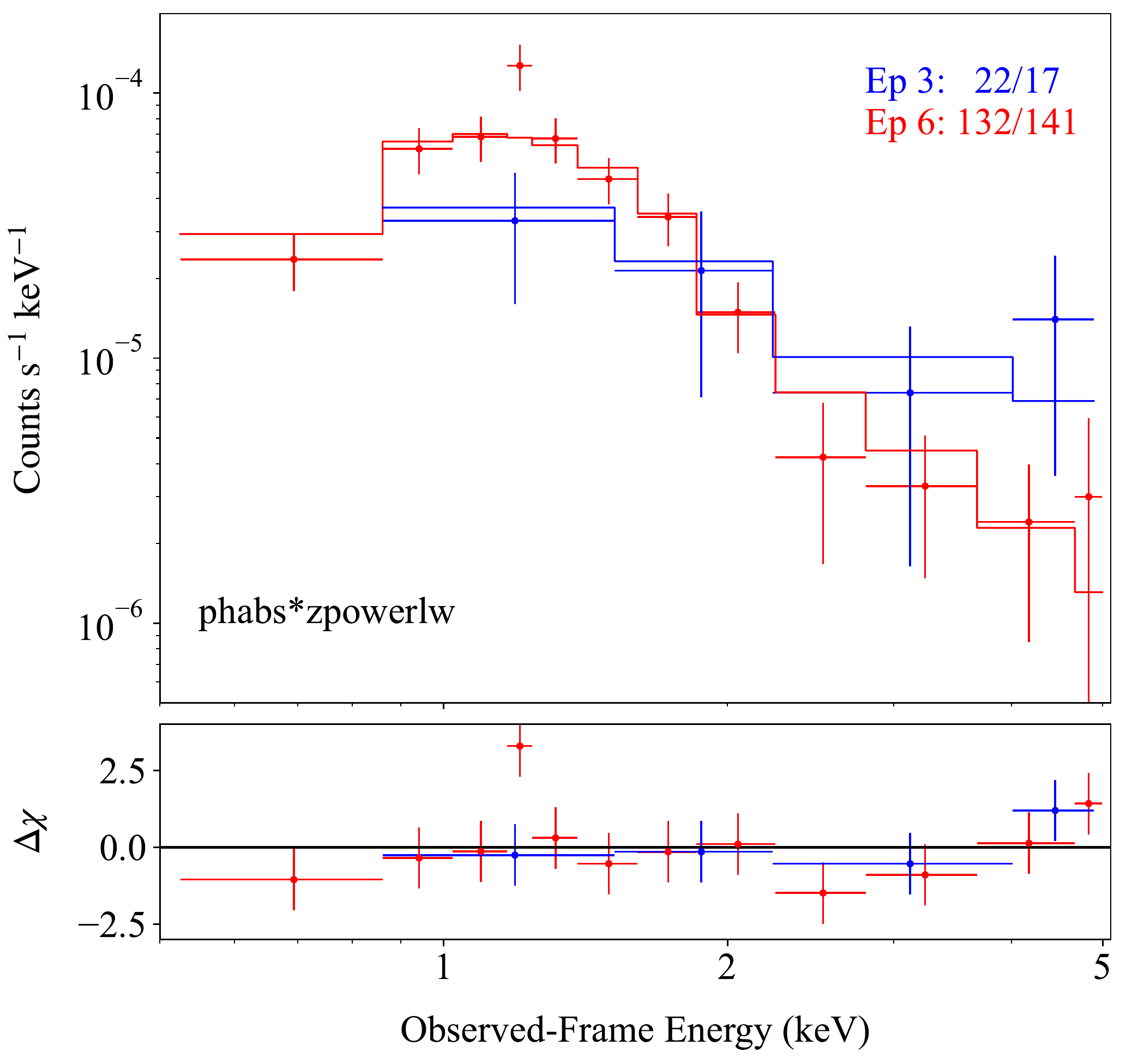}
\caption{The 0.5{--}5 keV X-ray spectra in epoch 3 (blue) and \hbox{epoch 6} (red), 
along with the \hbox{best-fit} \hbox{power-law} models. The model parameters are summarized in the first and second rows of Table~\ref{tab3}. The bottom panel shows  the fitting residuals. The spectra are grouped for display purposes only.  }
\label{xspec}
\end{figure}

\begin{figure*}
\centering
\includegraphics[width=3.1in]{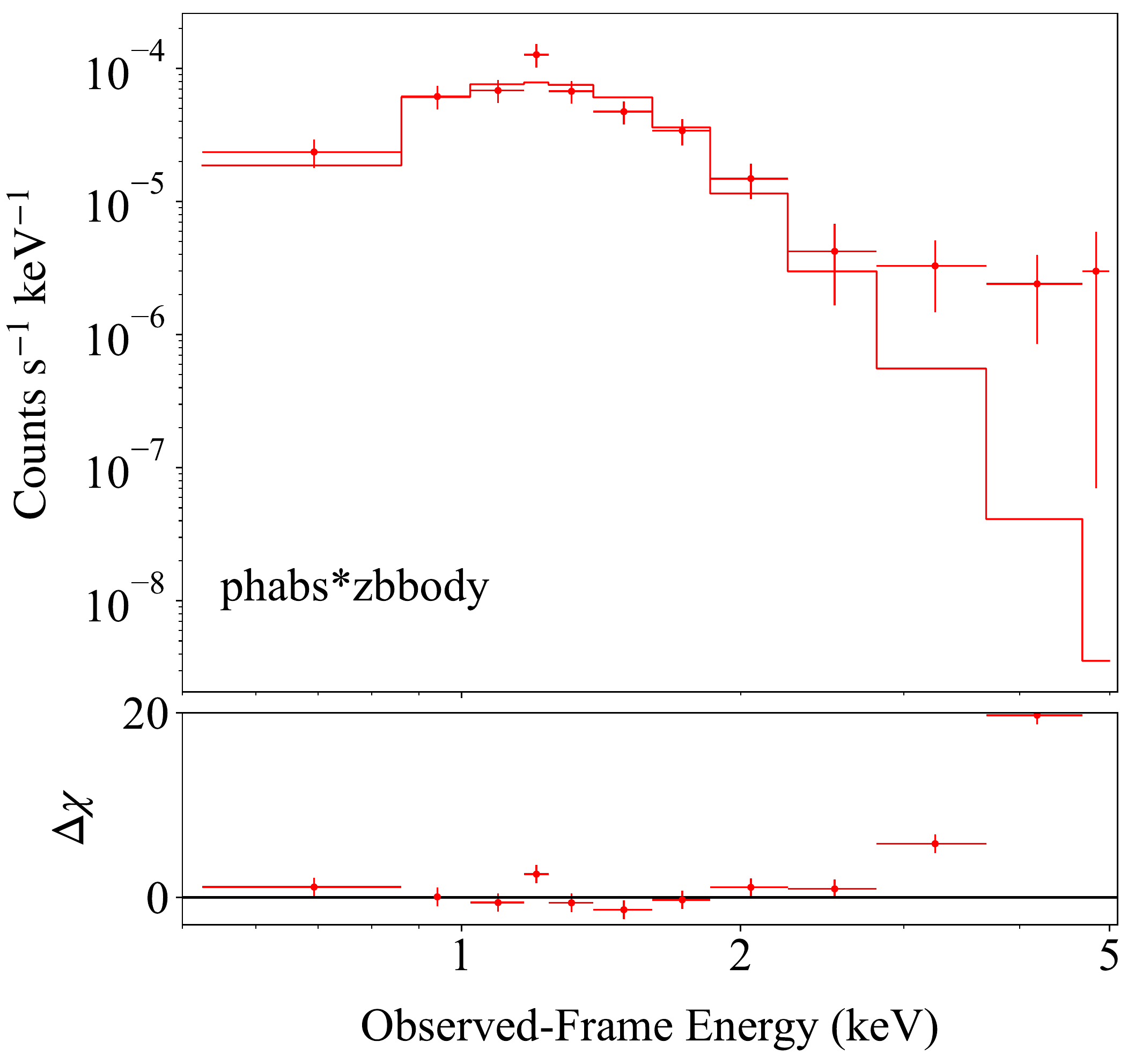}
\includegraphics[width=3.1in]{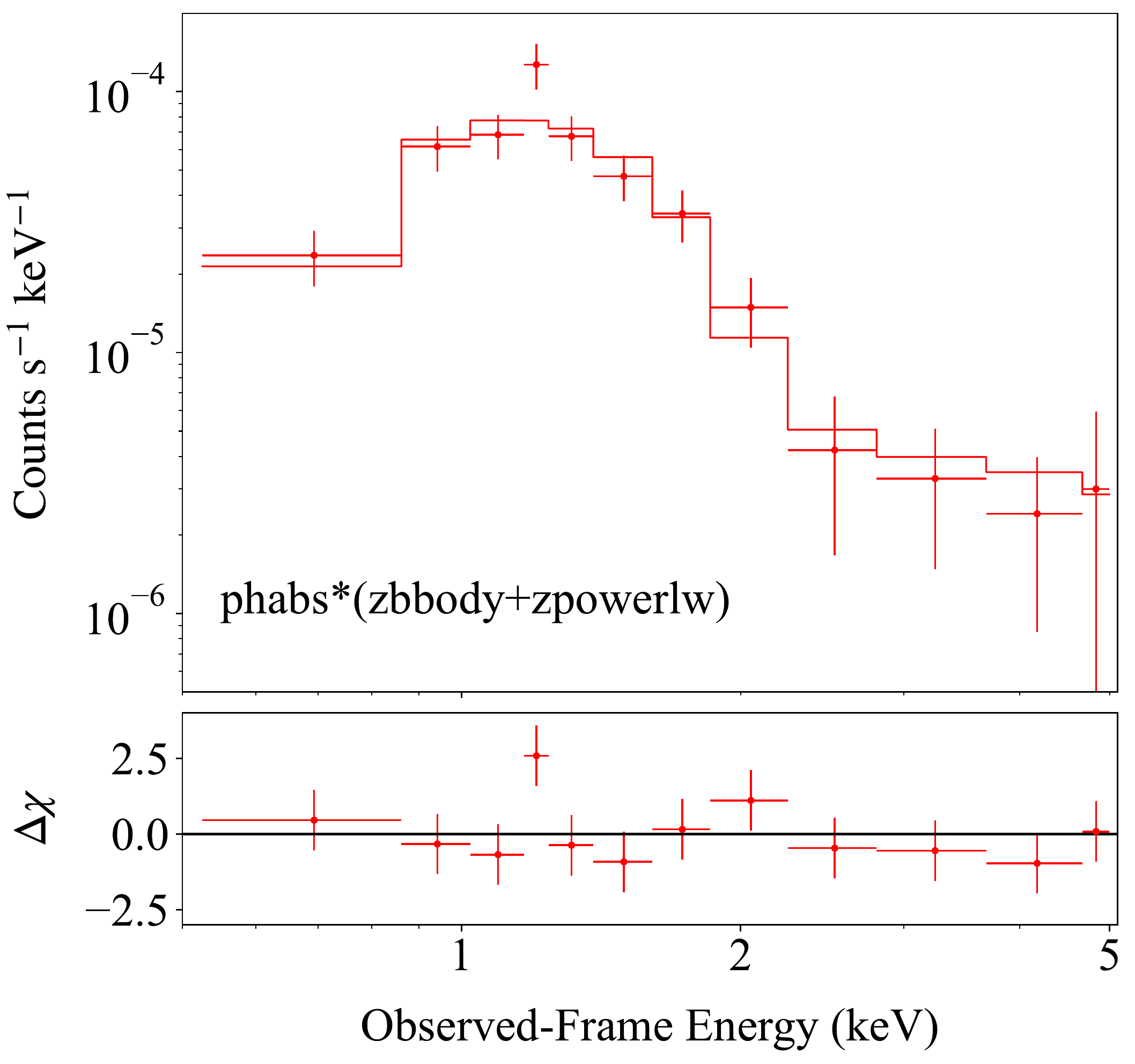}
\caption{
 The epoch 6 spectrum fitted with a black-body model (left panel) and a black-body plus power-law model (right panel). The models and the best-fit parameters are summarized in the third and fourth rows of Table~\ref{tab3}. The bottom panels show the fitting residuals. }
\label{zbbzpl}
\end{figure*}

\begin{deluxetable*}{clrrcccccr}
\label{tab3}
\tabletypesize{\scriptsize}
\tablewidth{0pt}
\tablecaption{Best-fit  Spectral Model Parameters}
\tablehead{
\colhead{Epoch}&
\colhead{Model}&
\colhead{$\Gamma$}&
\colhead{$\rm Norm_{\rm PL}$}&
\colhead{$kT$}&
\colhead{$\rm Norm_{\rm bb}$}&
\colhead{$N_{\rm H}$}&
\colhead{$\nu L_{\rm 2~\textup{keV}}$ }&
\colhead{ $L_{\rm X}$  }&
\colhead{$W/{\rm dof}$}\\
\colhead{Number}&
\colhead{}&
\colhead{}&
\colhead{ $(\times 10^{-6}) $}&
\colhead{(keV)}&
\colhead{ $(\times 10^{-8}) $}&
\colhead{ $(\times 10^{22}~\rm cm^{-2}) $}&
\colhead{$(\rm erg~s^{-1})$}&
\colhead{$(\rm erg~s^{-1})$}&
\colhead{}
}
\startdata
3 &phabs*zpowerlw& $1.2^{+0.7}_{-0.6}$ & $4.6^{+9.1}_{-3.1}$&--&--&-- &$3.5^{+6.9}_{-2.4} \times 10^{42}$&$1.2^{+2.4}_{-0.8} \times 10^{43}$&22/17   \\
\vspace{0.15cm}
6  &phabs*zpowerlw& $2.8\pm0.3$ &  $7.4_{-2.5}^{+3.9}$&--&--&-- &$1.7^{+0.9}_{-0.6} \times 10^{43}$ &$1.5^{+0.8}_{-0.5} \times 10^{43}$ & 132/141 \\
6  &phabs*zbbody& --& --&$0.73^{+0.06}_{-0.05}$&$8.6^{+0.8}_{-0.7}$&-- &$(1.1\pm0.1) \times 10^{43}$ &$(1.3\pm0.1)  \times 10^{43}$ & 142/141 \\
6  &phabs*(zbbody+zpowerlw)& 2.0 (fixed)& $1.1\pm0.3$&$0.60\pm0.08$&$6.2\pm1.3$&-- &$(1.4\pm0.3) \times 10^{43}$ &$(1.5\pm0.2) \times 10^{43}$ & 129/140 \\
3&phabs*zphabs*zpowerlw & $2.8$ (fixed) & 7.4 (fixed) &--&--& $5.0_{-2.4}^{+3.0}$ & -- & -- &24/18 \\
3&phabs*zpowerlw & $2.8$ (fixed) & $3.8^{+1.5}_{-1.3}$&-- &--&--&
$0.9^{+0.4}_{-0.3} \times 10^{43}$&$(0.8\pm0.3) \times 10^{43}$&25/18   \\
\enddata
%\tablenotetext{a}{The spectrum is fitted with an absorbed %power-law model (phabs*zphabs*zpowerlw)
%with $\Gamma$ and power-law normalization fixed at
%the epoch 6 values (see Section~\ref{sec4.3}).
%}
%\tablenotetext{b}{The spectrum is fitted with a simple power-law %model (phabs*zpowerlw) 
%with $\Gamma$ fixed at
%the epoch 6 value (see Section~\ref{sec4.3}).
%}
\end{deluxetable*}

We perform spectral analysis for the epoch 3 and \hbox{epoch 6} data using XSPEC (v12.11.1; \citealt{Arnaud1996}). The source and background spectra of individual observations  are extracted with AE, and they are then merged into the \hbox{epoch 3} and \hbox{epoch 6} spectra. 
The spectra are grouped using {\it grppha} so that each bin contains at least one count.
We use the XSPEC W statistic\footnote{\url{https://heasarc.gsfc.nasa.gov/docs/xanadu/xspec/manual/XSappendixStatistics.html}.}   for spectral fitting.
We  use a simple  \hbox{power-law} model modified by the Galactic absorption (phabs*zpowerlaw) to fit the \hbox{$\rm 0.5\textrm{--}5$ keV} spectra. The best-fit results and the derived luminosities are listed in Table \ref{tab3}. The \xray\ spectra with the best-fit models are displayed in Figure \ref{xspec}. The best-fit results are overall acceptable, considering the fitting statistics ($W/{\rm dof}$ in Table \ref{tab3}) and the fitting residuals. 

 In epoch 3, the spectral shape and normalization have large uncertainties due to the limited number of counts.
The best-fit $\Gamma$ is $1.2_{-0.6}^{+0.7}$; such a small $\Gamma$ value might indicate the presence of X-ray obscuration. The resulting $\nu L_{\rm 2~\textup{keV}}$ is $3.5^{+6.9}_{-2.4} \times 10^{42}~\rm erg~s^{-1}$, and the rest-frame \hbox{2{--}10~keV} luminosity ($L_{\rm X}$) is $1.2^{+2.4}_{-0.8} \times 10^{43}~\rm erg~s^{-1}$. The luminosity uncertainties are propagated from the uncertainty of the power-law normalization. 

In epoch 6, the best-fit photon index is $\Gamma =  2.8\pm0.3$ in the rest-frame 1.3--13 keV band. The resulting  $L_{\rm X}$ value is 
$1.5^{+0.8}_{-0.5}\times10^{43}~\rm erg~s^{-1}$. The best-fit $\Gamma$  is larger than those commonly observed in type 1 AGNs (a mean value of $\Gamma \approx 2.0$; e.g., \citealt{Scott2011,Liu2017}). It is even larger than those of typical super-Eddington accreting AGNs \citep[e.g.,][]  {Marlar2018,Huang2020}, suggesting a high accretion rate \citep[e.g.,][]{Shemmer2008}. 
The steep spectral shape also indicates that the \xray\ emission in epoch 6 is likely unobscured. 
Adding an intrinsic absorption component (zphabs) does not improve the fit, and we obtain an  upper limit of $2.5 \times 10^{22}~\rm cm^{-2}$ for the intrinsic $N_{\rm H}$. We also try to fit the epoch 6 spectrum with a cutoff power-law model (phabs*zcutoffpl) and examine if the photon index becomes smaller. The cutoff energy is loosely constrained. 
The $\Gamma$ value is still large ($\approx2.6$), even if we fix the cutoff energy to a small value of \hbox{15 keV}.

The \xray\ spectra of typical TDEs  can be described with a simple black-body model with  temperatures of  $\approx 10\textrm{--}100$~eV \citep[e.g.,][]{Saxton2021}.
We thus test a black-body model (phabs*zbbody) to fit the epoch 6 spectrum. The best-fit  temperature is $kT = 0.73^{+0.06}_{-0.05}$~keV with a \hbox{$W/{\rm dof}$} value of 142/141. The best-fit results are shown in Table~\ref{tab3} and the left panel of Figure~\ref{zbbzpl}.
This model describes well the observed-frame $\lesssim3$~keV spectrum, but there is significant excess emission above observed-frame $\gtrsim3$~keV energies. 
We then add an additional power-law component [phabs*(zbbody+zpowerlw)] to fit the high-energy excess. The best-fit results are overall acceptable ($W/{\rm dof} = 129/139$), with a
temperature of $0.65^{+0.07}_{-0.10}~\rm keV$ and $\Gamma$  of $0.5^{+2.2}_{-1.7}$. 
Since $\Gamma$ is loosely constrained, we fix it to 2.0,
and the best-fit temperature becomes \hbox{$0.60 \pm 0.08~\rm keV$}. The best-fit results are shown in Table~\ref{tab3} and the right panel of Figure~\ref{zbbzpl}.
Nevertheless, the best-fit temperatures ($kT\approx 0.60\textrm{--}0.73~\rm keV$) from these models are $\gtrsim 6$ times larger than  typical TDE temperatures (\hbox{$\approx 10\textrm{--}100$ eV}; \citealt{Saxton2021}). 

%But we caution that, given the rest-frame energy band of 1.3--13 keV, a lower temperature ($\lesssim 100~\rm eV$) black-body component cannot be constrained by the epoch 6 \xray\ spectrum. 
%We do not consider that such a component would emerge in a softer energy band, because the \xray\ luminosity would be several orders of magnitude larger.

We also break the epoch 6 spectrum into two segments to investigate if there is any spectral evolution. The first segment consists of  the first three subepochs and the second segment consists of the last three subepochs.
 We fit the two spectra with the simple power-law model. The best-fit parameters are \hbox{$\Gamma = 2.9\pm0.4$} and \hbox{norm = $8.4^{+5.8}_{-3.4} \times 10^{-6}$} for segment 1 with \hbox{$W/\rm dof = 69/100$}, and they are \hbox{$\Gamma = 2.8^{+0.5}_{-0.4}$} and \hbox{norm = $6.7^{+3.1}_{-5.8} \times 10^{-6}$} for segment 2 with \hbox{$W/\rm dof = 93/113$}. Thus there is no apparent spectral evolution. This result is consistent with the lack of variability determined from  the photometric analysis in Section \ref{sec:LC}.

\subsection{Optical Variability and R-band Light Curve}\label{sec2.2}

\begin{figure}
\includegraphics[width=3.3in]{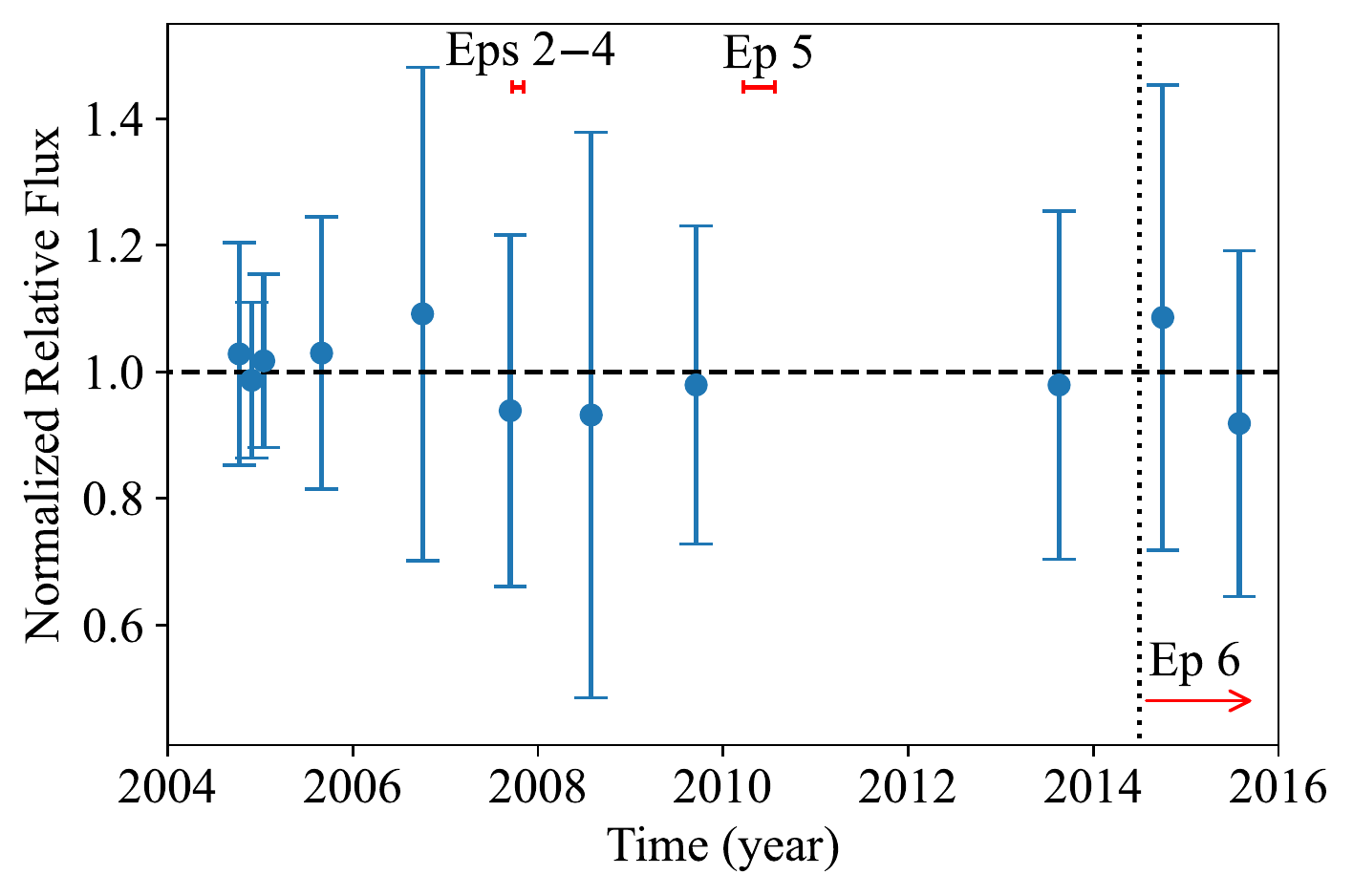}
\caption{VLT/VIMOS \rband\ light curve of XID 403. The data points show the  relative flux measurements grouped  within one month, and they are normalized to their weighted average.  
The two horizontal red bars denote the observation date ranges for \xray\ epoch 2--4 and \hbox{epoch 5}. The vertical dotted line indicates the start date of epoch 6.}
\label{oplc}
\end{figure}

%Using the combined \xray\ data  with a total exposure time of 1.06 Ms from June 2014 to October 2014, \cite{Luo2014} reported the \xray\ flux increase of \xid. They did not find any contemporaneous \rband\ flux variability.
We first search several multi-epoch optical photometric catalogs for any optical variability of \xid, including the  PAN-STARRS1 (DR2; \citealt{Chambers2016}), the Dark Energy Survey (DES DR2; \citealt{DES2021}), the Zwicky Transient Facility (ZTF DR9; \citealt{Masci2019}),  the Catalina Real-Time Transient Survey (CRTS DR2; \citealt{Drake2012}), and the  Hubble Catalog of Variables (HCV; \citealt{Bonanos2019}).
\xid\ is only present in the HCV. The HCV light curves of \xid\ cover a period from  2002 to 2012, and they show only mild potential  optical variability in the $V_{660}$, $i_{775}$, $z_{850}$, and F814W bands. We apply the \cite{Villforth2010} $\chi^2$ method (similar to the  $\chi^2$ method used  for the \xray\ data in Section \ref{sec:LC}) and investigate if we can identify \xid\ as an AGN based on the variability.
The resulting $\chi^2$ values and the corresponding null-hypothesis probabilities (ranging from $\approx 25\%\textrm{--}87\%$) for the HCV light curves are not sufficiently high to meet the AGN selection criterion (e.g., $>99.9\%$ as in \citealt{Villforth2010}). Thus, the optical variability in the HCV light curves is not significant.

The CDF-S was also monitored by  OmegaCAM \citep{Kuijken2011} on the VLT Survey
Telescope (VST) in the $u$, $g$, $r$, and $i$ 
bands. \cite{Falocco2015} used these data to perform variability selection of AGNs. \xid\ was not included in this study due to its low $r$-band flux.
We reduce and process the $r$-band data ranging from 2012 to 2013,  following the same  method in \cite{DeCicco2019}, and there is no significant variability in the resulting light curve.

%We  search the ESO archive for the VST/OmegaCAM data.
%The VST/OmegaCAM consists of 32 CCDs with a  field-of-view (FoV) of $1\deg \times 1\deg$, so we require the pointing position of the imaging observation to be within $1\deg$ of \xid. We also require the exposure time to be greater than 100 s.  Compared to the other bands, the  $r$-band exposure times are the longest. The central wavelength of the $r$ band is close to that of the $R$ band, and thus we use only the $r$-band data.
%We only use the $r$-band data because of the longest exposure time compared to the other bands and the similarity with $R$ band. 
%In the end, we obtain 247 \hbox{$r$-band} images with observation dates ranging from July 2012 to March 2018. The exposure time of each image is either \hbox{360 s} or \hbox{300 s}. We reduce and process the $r$-band data following the same method described in \cite{DeCicco2019},
%and there is still no sign of strong optical variability.

Since the HCV and VST light curves do not overlap with the epoch 6 observations, we further search the European Southern Observatory (ESO) archive\footnote{\url{http://archive.eso.org/cms/eso-data.html}.} for imaging data to construct a longer-term light curve.
 %for archival optical imaging data. 
To obtain reliable photometric measurements, we require the pointing position of the imaging observation to be within 15\arcmin\ of the \xid\ position and the exposure time to be larger than 100 s. We also require the optical band to have imaging data after June 2014, the start of the epoch 6 observations.
 Only 189 $R$ band images obtained by the Very Large Telescope using the Visible MultiObject Spectrograph (VIMOS; \citealt{LeFevre2003}) satisfy this requirement. The observation dates  range from November 2003  to July 2015.
%We also match the position of \xid\ to Pan-STARRS1 DR2 \citep{Chambers2016} and Dark Energy Survey (DES) DR2 \citep{DES2021} archives, and no any matches are found.

 To construct an \rband\ light curve from the VLT/VIMOS \rband\ images, we perform aperture photometry using the Python package {\sc photutils} \citep{photutlis}. A nearby star J033223.18$-$274139.1 (hereafter J0332; blue square in Figure \ref{opim}), which is 6.6\arcsec\ to the northeast from XID 403, is used for flux calibration. 
 The  star is 23.4 times brighter than \xid\ in the $R$ band \citep{Straatman2016}, and based on its HCV data, it is not variable.
We perform source detection in the \hbox{$20\arcsec \times 20\arcsec$} region centered on the star J0332, adopting a \hbox{2.0\arcsec-radius} circular aperture and a $3\sigma$ detection threshold.
XID 403 is detected in 87 of the 189 images.  The other images typically  have short exposure times;  the upper limits on the source fluxes from these images are not constraining compared to the measurements, and thus we do not use these images in the following analysis.
Besides XID 403 and J0332, there are typically 1{--}4 sources  detected in the source searching region.
One of the sources is the nearby  brighter galaxy, which is not variable either based on the HCV data.

To obtain the \rband\ flux of XID 403 in each image, we first determine the background photon counts per pixel from the \hbox{$20\arcsec \times 20\arcsec$}  region described above. All detected sources   are masked out  with a \hbox{4.0\arcsec-radius circular aperture}.
We subtract the background from the image to produce a background-subtracted image. Then we extract the net source counts of \xid\  and J0332  with a \hbox{1\arcsec-radius} circular aperture  and a \hbox{1.5\arcsec-radius} circular aperture, respectively. A smaller aperture is used for \xid\ to reduce contamination from the nearby galaxy.\footnote{One caveat is that the source is slightly extended in the {\it HST} images (e.g., see Figure \ref{opim}), and the extension might also depend upon seeing for VLT/VIMOS observations. Thus 
the \hbox{1\arcsec-radius} circular aperture likely underestimates the total galaxy flux and might also cause seeing-dependent fluctuations. However, since we are searching for AGN related strong variability and we do not find any in the end, this approach appears sufficiently robust for our purposes.}
Since the star J0332 is not variable, we calculate the relative flux of \xid\ as the ratio of its net source counts to that of J0332. 
The 1$\sigma$ uncertainties of the relative fluxes are calculated by propagating  the 1$\sigma$ uncertainties of the source counts.

To assess the contamination from the nearby brighter galaxy in the above extraction, we estimate the fraction of counts in the 1\arcsec-radius extraction aperture that is likely from this galaxy.  We use an image of the star J0332 to approximate the point spread function (PSF) of the VLT/VIMOS imaging. The images of the $z = 0.535$ brighter galaxy do not show any clear extension, and we also verify that it does not vary between the observations. Thus we simulate two point sources with a separation of 1.9\arcsec\ and a flux ratio of 5.7. The contamination (fraction of counts) from the brighter source in the 1\arcsec-radius aperture of the fainter source is $\approx8\%$. Thus we consider that the nearby brighter galaxy does not affect the derived relative fluxes significantly, nor does it affect our assessment of the $R$-band variability.

Using the derived relative fluxes, we construct an \rband\ light curve for \xid. There are 87 flux measurements. 
We group flux measurements within one month, and adopt the weighted average of the relative flux for each group. The 1$\sigma$ uncertainty of each grouped flux is the combination of the weighted average of  measurement uncertainties and the weighted standard deviation of the measurements.
The light curve is shown in Figure \ref{oplc}, including 11 grouped data points normalized to their weighted average.
There are two data points within epoch 6, and the other points spread in the gaps between the epochs. Considering the  uncertainties, there is no significant \rband\ variability, and there is no \rband\ brightening contemporaneous to the \xray\ brightening in epoch 6.
We also use the $\chi^2$ method \citep{Villforth2010} to assess the significance of the variability, and the null hypothesis probability is very small ($\approx 0.0001$), indicating that it is not variable.

Since there is no detectable variability, we place an upper limit on the $R$-band brightening factor in epoch~6. We assume that the $R$-band fluxes prior to epoch~5 are dominated by the host galaxy, and we group all these data points (the left eight data points) in Figure~\ref{oplc} and compared their average relative flux ($1.00\pm0.12$) to that of the two data points after the start date of epoch~6 ($0.98\pm0.24$). We simulate the two average relative fluxes using the Monte Carlo approach described in Section~\ref{sec:LC}, assuming they follow Gaussian distributions. The resulting 90\% confidence-level upper limit on the variability factor is 1.33. Therefore, if there is an optical/UV flare contemporaneous to the epoch~6 \xray\ outburst from an AGN or TDE, its $R$-band flux is $< 33\%$ of that from the host galaxy.

\subsection{Spectral Energy Distribution }\label{secSED}

\begin{figure*}
\centerline{
\includegraphics[width=6.0in]{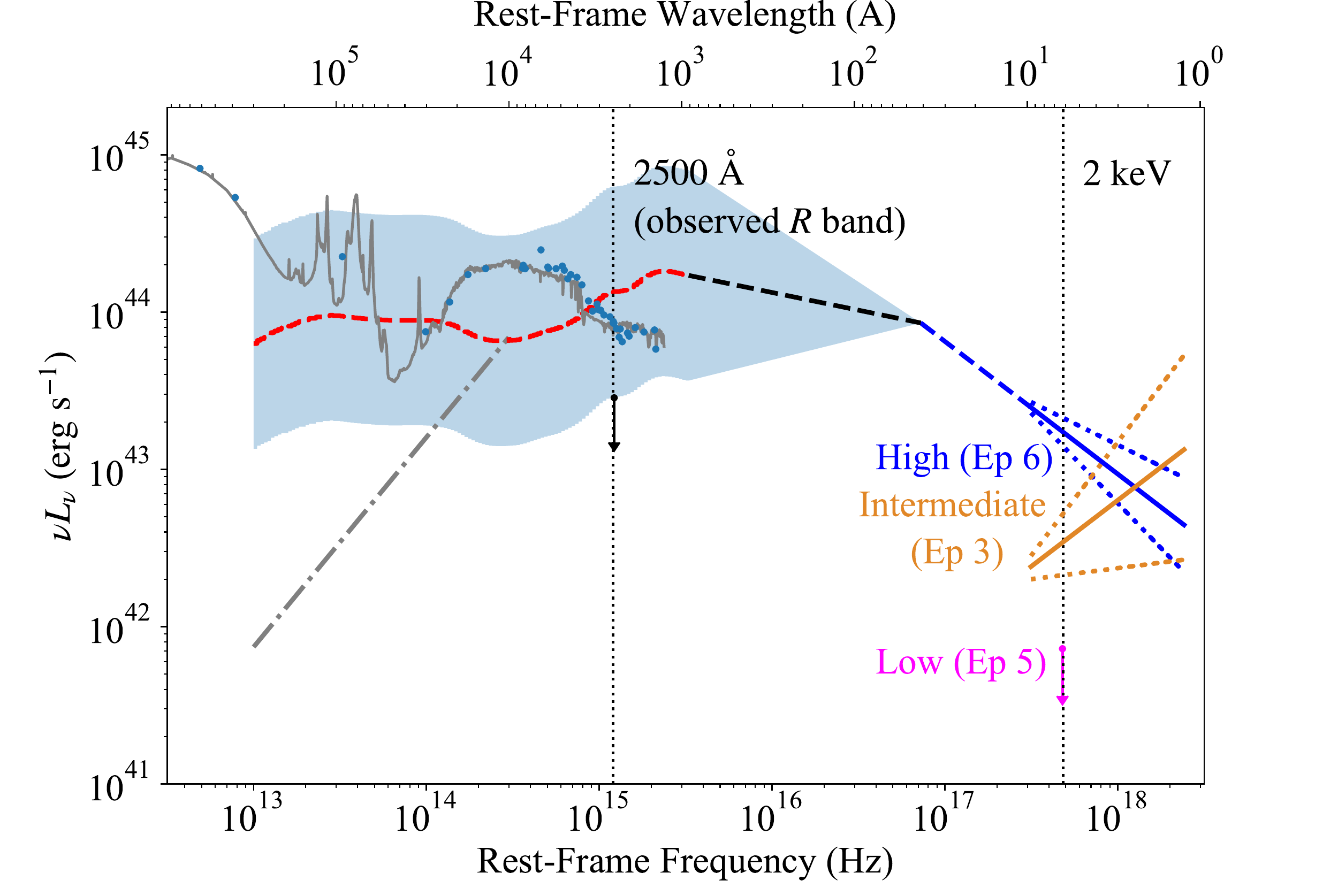}}
\caption{Rest-frame SED of XID 403. The blue points are the photometric data from IR to UV \citep{Straatman2016}.  
The gray curve is the {\sc cigale} best-fit SED model, and the line emission from the nebular component is not shown for display purpose.
The blue solid line is the high--state (epoch 6) 1.3{--}10 keV \xray\ SED from the spectral fitting results (see Section \ref{sec3.2}), and the blue dotted lines show the uncertainties of the photon index. Similarly, the dark orange solid and dotted lines indicate the epoch~3 power-law spectrum and the uncertainties of the photon index. 
The blue dashed line is
the 0.3{--}1.3 keV band \xray\ SED extrapolated from the high-state spectral fitting results. 
The magenta arrow is the $\nu L_{\rm 2~\textup{keV}}$ upper limit from epoch 5 for the low state. The black downward arrow indicates the upper limit on the $R$-band luminosity contributed by any flaring component in epoch 6 inferred from the $R$-band light curve in Section~\ref{sec2.2}.
The red dashed curve (from \hbox{30 $\mu$m} to 912 \AA) is the low-luminosity quasar UV-to-IR SED template \citep{Krawczyk2013} scaled to the 2500 \AA\ luminosity  derived from the \cite{Steffen2006} \al\ relation  and the epoch~6 spectrum. 
The black dashed line (from 912 \AA\ to 0.3 keV) is a power law connecting the optical/UV SED template and the \xray\ SED.
The blue shaded region illustrates the allowed range of the expected AGN UV-to-IR SED due to the large scatter of the \al\ relation.
The gray dash-dotted line (from 30 $\mu$m to 1 $\mu$m) is an \hbox{$\alpha_\nu$ = 1/3} power law, representing the IR 
SED from the accretion disk. 
The two vertical dotted lines denote the locations of \hbox{2500 \AA} (observed $R$ band) and \hbox{2 keV}.
}
\label{sed}
\end{figure*}

The CDF-S has superb multiwavelength coverage, allowing construction of a broad band spectral energy distribution (SED) for \xid. We adopt the UV-to-infrared (IR) photometric data from \cite{Straatman2016},
which include \hbox{41 bands} ranging from 3686 \AA\ to 160 $\mu$m and have been corrected for Galactic extinction. 
XID 403 was not detected in the sensitive Very Large
Arrary (VLA) 1.4 GHz survey of the CDF-S \citep{Luo2017}.
We plot the SED in \hbox{Figure \ref{sed}}. 
There is no apparent AGN component in the optical/UV SED.  The strong \hbox{far-IR} (FIR) radiation suggests a high \hbox{star-formation} rate of the host galaxy. 
All the observation dates of the UV-to-IR photometric data are before \hbox{epoch 6}, and thus the SED in \hbox{Figure \ref{sed}} does not necessarily represent the SED of \xid\ during its high \xray\ state (epoch 6). However, given its stable \rband\ (\hbox{rest-frame 2461} \AA) light curve (Figure \ref{oplc}), we do not consider that an AGN component would have emerged in its high-state SED.

We also add \xray\ constraints to the SED. 
For \hbox{epochs 3 and 6}, we plot in Figure \ref{sed} the best-fit power-law spectra (dark orange and blue solid lines) in the rest-frame 1.3--10 keV band,
%with the cyan and blue solid lines in a rest-frame energy band of 1.3--10 keV. 
where the rest-frame \hbox{1.3 keV} corresponds to the observed-frame 0.5 keV. The $1\sigma$ uncertainties for $\Gamma$  are represented with the corresponding dotted lines.
For the  low states, epoch 5 has the longest exposure, providing the most stringent upper-limit constraint on the low-state luminosities. We convert the upper limit on the full-band counts  to a  \hbox{2 keV} flux upper limit using the spectral response files and a $\Gamma = 2$ power-law spectrum. The derived $\nu L_{\rm 2~\textup{keV}}$ upper limit  is $7.2 \times 10^{41}~\ergs$ and  it is shown as the gray arrow in \hbox{Figure \ref{sed}}. 

To obtain  basic properties of the host galaxy, including its stellar mass ($M_{\ast}$) and star-formation rate (SFR),
we perform SED fitting using the Python package {\sc cigale} \citep[e.g.,][]{Boquien2019,Yang2020,Yang2022}.
A set of SED templates are generated based on five modules, including a delayed star-formation history with an optional exponential burst component (\hbox{sfhdelayed}), a library of simple stellar populations (bc03; \citealt{bc03}), a nebular emission component (nebular), a dust attenuation law with $E(B-V)$ values ranging from 0 to 0.4 mag (\hbox{dustatt\_calzleit}; \citealt{Calzetti2000}), a dust emission component (dale2014; \citealt{Dale2014}), and a fixed redshifting. The \cite{Chabrier2003} stellar initial mass function (IMF) is used. The best-fit SED model, shown in Figure \ref{sed},  describes well the FIR-to-UV SED of XID 403 with a reduced $\chi^2_r$ value of 0.95. An AGN component is not required in the above SED fitting, and adding such a component (skirtor2016; \citealt{Stalevski2012,Stalevski2016}) does not improve the fit.

From the best-fit model, we derive stellar mass and SFR values of $(2.7 \pm 0.2) \times 10^{10}~\msun$ and \hbox{$(46 \pm 3)~\msun~\rm yr^{-1}$}, respectively.
Considering that there may be additional systematic uncertainties, our stellar mass and SFR measurements are likely consistent with those in \cite{Trump2013} and \cite{Straatman2016}. 
The specific SFR (${\rm sSFR}={\rm SFR}/M_{\ast}$) of \xid\ is  $1.7~\rm Gyr^{-1}$, within the ``main-sequence'' of star-forming galaxies at the same redshift (e.g., \citealt{Elbaz2011}).

 \subsection{Estimations of AGN Bolometric Luminosity and SMBH Mass}\label{sec3.4}
 %To estimate the epoch 6 AGN 
Under the assumption that XID 403 is an AGN, 
 we estimate the AGN bolometric luminosity in \hbox{epoch 6}  by constructing an expected intrinsic AGN SED based on the best-fit \xray\ spectrum, which is considered to be the intrinsic \xray\ emission (i.e., without \xray\ obscuration) given the very steep spectral shape.  In this case, XID 403 should be a type 1 AGN at least in epoch 6 as there is no X-ray obscuration.
 We extrapolate the \xray\ SED from \hbox{1.3--10 keV} to \hbox{0.3{--}10 keV} (Figure \ref{sed}). 
 We then infer a 2500 \AA\ monochromatic luminosity ($L_{\rm 2500~\textup{\AA}}$) from  $\nu L_{\rm 2~\textup{keV}}$ and the \cite{Steffen2006} $\al$ relation (Equation 2), which is \hbox{$1.1^{+4.1}_{-0.9} \times 10^{29}~\ergs~\rm Hz^{-1}$}.
The uncertainties of  $L_{\rm 2500~\textup{\AA}}$ are propagated from the rms scatter of the expected $\alpha_{\rm OX}$ value (0.165; see Table 5 of \citealt{Steffen2006}).  
%The  ratio of the expected AGN $L_{\rm 2500~\textup{\AA}}$ to the 2500 \AA\ host luminosity is  $1.5^{+5.6}_{-1.2}$.
 For the UV-to-IR SED (30 $\mu$m to 912 \AA), we adopt the low-luminosity quasar SED template (given the epoch 6 \xray\ luminosity) in \cite{Krawczyk2013}, and scale it to $L_{\rm 2500~\textup{\AA}}$. The  912~\AA\ to 0.3 keV SED is a power law connecting the UV and the \xray\ SEDs. 
The expected intrinsic SED and its allowed scatter are shown in Figure \ref{sed}; the scatter is computed from the uncertainties of $L_{\rm 2500~\textup{\AA}}$. Considering the scatter of the expected AGN SED, there is a broad range for the  ratios of AGN luminosities to the host luminosities (e.g., $1.5^{+5.6}_{-1.2}$ at 2500 \AA). 
To avoid double counting the IR emission that is mostly reprocessed emission from the dusty torus, we extrapolate a power-law SED with a spectral slope of $\alpha_\nu = 1/3$ \citep[e.g.,][]{Davis2011,Liu2021} from 1 $\mu$m to 30 $\mu$m (gray dash-dotted line in Figure \ref{sed}). We integrate the 30 $\mu$m-to-10 keV AGN SED and obtain a  $L_{\rm bol}$ value of $7.8^{+18.2}_{-3.9} \times~10^{44}~\ergs$. 
The uncertainty is dominated by the uncertainty of $L_{\rm 2500~\textup{\AA}}$.
 Compared to  typical AGNs, super-Eddington accreting AGNs are expected to emit excess EUV radiation \citep[e.g.,][]{CN2016,KD2018}. Thus there may be additional uncertainties on $L_{\rm bol}$ from the above SED integration if \xid\ is indeed super-Eddington accreting given its steep spectral shape in epoch 6.
 
 We note that the above estimation is based solely on the epoch 6 \xray\ spectrum and an empirical understanding of typical AGN SEDs (including the $\al$ relation), without invoking constraints from optical observations. On the other hand, from the $R$-band light curve presented in Section~\ref{sec2.2}, we inferred that if there is AGN emission emerging in epoch~6, its contribution to the $R$-band flux is $<33\%$. This upper limit constraint is illustrated by the black downward arrow in Figure~\ref{sed}, which is slightly below the allowed scatter of the expected AGN SED (blue shaded region). Using this upper-limit value to normalize a typical AGN SED, we would obtain an upper limit on the AGN bolometric luminosity of $<3.4\times10^{44}~\rm erg~s^{-1}$. This discrepancy indicates that the epoch~6 outburst is unlikely due to the emergence of a type~1 AGN with a standard broad-band SED. A simple explanation is dust reddening; mild dust extinction is able to suppress efficiently  the observed $R$-band (rest-frame $\approx2500$~\AA) flux and UV SED reddening is not uncommon in type~1 AGNs or luminous quasars. We present quantitative discussion of the possible dust extinction in Section~\ref{sec4.3} below. 
 %Moreover, reddening has also been invoked to explain \xray\ TDEs without optical/UV counterparts \citep[e.g.,][]{Burrows2011,Gezari2021}.
 Therefore, considering the reddening effects, the stable $R$-band light curve in Section~\ref{sec2.2} might not be very constraining for the nature of the 
\xray\ outbursts.

We estimate the SMBH mass ($\mbh$) of XID 403 from its stellar mass ($2.2 \times 10^{10}~\msun$) using the \hbox{$\mbh\textrm{--}M_{\ast}$} scaling relation for AGNs (Equation 5 of \citealt{Reines2015}). The resulting SMBH mass is $ 5 \times 10^6~\msun$.\footnote{The  \cite{Chabrier2003} IMF was used consistently when deriving the stellar mass of XID 403 and the \cite{Reines2015} $\mbh\textrm{--}M_{\ast}$ scaling relation.} 
The uncertainty of this $\mbh$ value is estimated to be \hbox{$\approx 0.56$ dex},
dominated by the scatter of the scaling relation  (0.55 dex; see \hbox{Section 4.1} of \citealt{Reines2015}). Since there are additional uncertainties ($\approx 0.5$ dex; e.g., \citealt{Shen2013}) in the single-epoch virial SMBH mass estimates used to derive the \cite{Reines2015} relation,
 the $\mbh$ uncertainty of \xid\  could be even larger.
 We note that the $\mbh$ uncertainty does not affect significantly our following discussion. For example, if  $\mbh$ is larger by an order of magnitude, the Eddington ratio estimates would be an order of magnitude smaller (which are very uncertain anyway), and the size and variability time constraints in Section \ref{sec:discussion} below would vary, but our main conclusions (i.e., two \xray\ unveiling events; Section~\ref{sec4.3}) remain the same.

%slightly larger than the typical uncertainty of the virial SMBH masses, i.e., 0.5 dex \citep{Shen2011}.
%The uncertainty of this $\mbh$ value is estimated to be \hbox{$\approx 0.25$ dex}, which is the combination of the intrinsic scatter (\hbox{$\approx0.24$ dex}; \citealt{Reines2015}) of the scaling relation and the uncertainty of the stellar mass. We note that this uncertainty might be underestimated, because the typical uncertainty of the virial SMBH masses, which were used to calibrate the scaling relation, is estimated to be 0.5 dex \citep{Shen2011}. If we add this value into the uncertainty estimation, the $\mbh$ uncertainty would be \hbox{$\approx0.56$ dex}.
%We estimate the SMBH mass ($\mbh$) of XID 403 from its stellar mass ($2 \times 10^{10}~\msun$; \citealt{Trump2013}) using the \hbox{$\mbh\textrm{--}M_{\ast}$} scaling relation for AGNs (Equation 5 of \citealt{Reines2015}).
%The resulting SMBH mass is $5 \times 10^6~\msun$. 

We then estimate the \hbox{epoch 6} Eddington ratio to be \hbox{$\lambda_{\rm Edd} = L_{\rm bol}/L_{\rm Edd} = 1.2^{+2.9}_{-0.7}$}.  This Eddington ratio is consistent with the  value ($1.3^{+4.9}_{-1.0}$) derived from the \hbox{$\lambda_{\rm Edd} \textrm{--} \Gamma$} relation \citep[Equation 2 of ][]{Shemmer2008}, given its $\Gamma$ value of  $2.8\pm0.3$. 
The above practice can also be applied to epoch 3, but its flat spectral shape ($\Gamma=1.2^{+0.7}_{-0.6}$) suggests that the observed \xray\ spectrum is modified by intrinsic absorption. We thus add an absorption component (zphabs) to fit the epoch~3 spectrum and derive the absorption-corrected continuum. Since the spectrum has a limited number of counts, we fix the power-law photon index to a typical value of 2.0. The resulting \nh\ value is $2.3^{+4.9}_{-2.3} \times 10^{22}~\rm cm^{-2}$. 
We adopt the same SED integration method and  derive a $L_{\rm bol}$ value of $\approx 2.2 \times~10^{44}~\ergs$, and the corresponding Eddington ratio is $\approx 0.33$.

We also estimate  $L_{\rm bol}$ and $\lambda_{\rm Edd}$ for epochs 1, 2, 4, and 5 from their \xray\ upper limits, again based on the \al\ relation and typical AGN SEDs. A $\Gamma = 2.0$ power-law \xray\ spectrum is adopted. We convert the upper limits on the full-band counts to 2 keV flux upper limits using the spectral response files (with no absorption correction). The resulting upper limits on the  2 keV luminosities for epochs 1, 2, 4, and 5 are 0.80, 4.0, 0.99, and $0.72 \times 10^{42}~\rm erg~s^{-1}$, respectively. We notice that the full-band PF of epoch 6 is higher than that of epoch 5 by a factor of $> 12.6$, but the \hbox{2 keV} luminosity  of epoch 6 is higher than that of epoch 5 by a factor of $> 23.6$. This is the consequence of the  steeper \xray\ spectral shape used in determining the \hbox{2 keV} luminosity of \hbox{epoch 6.}
Using the same SED integration method, the resulting upper limits on the  bolometric luminosities for epochs 1, 2, 4, and 5 are 9.1, 76, 12, and 7.9 $\times 10^{42}~\rm erg~s^{-1}$, respectively. The corresponding upper limits on the Eddington ratios are 0.014, 0.12, 0.018, and 0.012,  respectively. We caution that these estimates are based on the assumption that the $\nu L_{\rm 2~\textup{keV}}$ upper limits are intrinsic, but these non-detections might instead be due to \xray\ absorption. Thus the Eddington ratios could be much higher. The difference ($\approx 100$) in Eddington ratios between  epoch 5 and epoch 6 is much larger than the difference ($\approx 12$) in their \xray\ fluxes, mainly due to the different $\Gamma$ values used and the  non-linear relation between $L_{\rm 2~\textup{keV}}$ and $L_{2500~\textup{\AA}}$.

\section{Discussion}\label{sec:discussion}
The main properties of CDF-S XID 403 and its extreme \xray\ variability are summarized as follows.
\begin{enumerate}
\item There were two \xray\ brightening events. 
The epoch 3 outburst (intermediate state; detected at a 3$\sigma$ significance level) appears to evolve quite fast. Given the epoch 2--4 light curve, \xid\ brightened by a factor of $>2.5$ in $\lesssim6.1$ rest-frame days, the outburst lasted for \hbox{$\approx5.0$--7.3 days}, and then the source dimmed by a factor of $>6.0$ in $\lesssim6.1$ days.
The \hbox{epoch 6}  outburst (high state) happened $\approx1.1\textrm{--}2.5$ years later in the rest frame. It lasted over $>251$ days and has a large variability amplitude (e.g., $>12.6$). There is no significant spectral evolution within epoch~6.

\item The epoch~3 spectrum is described with a $\Gamma=1.2^{+0.7}_{-0.6}$ power-law, suggestive of \xray\ absorption. 
The epoch~6 \xray\ spectrum in the rest-frame \hbox{1.3--13 keV}  band is described with a $\Gamma = 2.8\pm0.3$ power law. 
%There is no obscuration signature, and the 2 keV luminosity reaches $1.7^{+0.9}_{-0.6} \times 10^{43}~\rm erg~s^{-1}$.

\item 
The observed FIR--UV SED of \xid\ is dominated
by the host galaxy, which has a stellar mass of \hbox{$\approx 2.7 \times 10^{10}~\msun$} and a star-formation
rate of \hbox{$\approx 46~\msun~\rm yr^{-1}$}.
There is no significant optical/UV variability and 
there is no
$R$-band (\hbox{rest-frame} \hbox{$\approx2500$~\AA}) 
brightening ($<33\%$) contemporaneous with the epoch 6 \xray\ brightening.

\item There is no clear AGN signature in
the available NIR spectra or the IR-to-UV SED, but \xid\ should be either an AGN or associated with TDEs given the luminous \xray\ emission. 

\end{enumerate}

The discovery of \xid\ involves unique data sets; i.e., \hbox{7 Ms} of \chandra\ exposure. 
%We thus do not expect any precedents of such extreme \xray\ events at high redshifts.  However, it is somewhat surprising that we do not find any of its local analogs in the literature. 
We thus
do not expect any precedents of such extreme \xray\ variability events  at
high redshifts, which require multiple $\gtrsim100$~ks 
\chandra\ exposures. However, we do
not find any close local analogs of XID 403
either.
AGNs with \xray\ variability events induced by changes of accretion rate usually show changes in  optical properties \citep[e.g.,][]{Gilli2000,LaMassa2015,Dexter2019, Ai2020,Wang2020,Lyu2021}. %TDEs are generally identified in inactive galaxy \citep[e.g.,][]{Komossa2015,Gezari2021}.  
Typical TDEs do not show  recurrent outbursts \citep[e.g.,][]{Zabludoff2021} or steep \xray\ spectral shapes up to $\approx$ 13 keV \citep[e.g.,][]{Saxton2021}.
Type 2 AGNs with changes of \xray\ obscuration do not appear to recover to \xray\ unobscured  states \citep[e.g.,][]{Matt03,Guainazzi2002,Bianchi2005,Rivers2011,Rivers2015b,Marchese2012,Braito2013,Marinucci2016}. 
The \xray\ variability of \xid\ is reminiscent of local narrow-line Seyfert 1 galaxies (NLS1s) that sometimes show strong \xray\ variability due to variable partial-covering absorption \citep[e.g.,][]{Tanaka2004,Turner2009,Miniutti2012,Liu2021} and/or disk reflection \citep[e.g.,][]{Fabian2004,Fabian2012,Grupe2007,Grupe2008,Parker2014,Grupe2019,Parker2019b}, but unlike NLS1s, its NIR spectra (Section \ref{prop}) and SED (Section \ref{secSED}) show no AGN  signature.
 We explore in more detail each of the scenarios in the following subsections.

\subsection{Change of Accretion Rate?}\label{sec4.1}

In this scenario, the \xray\ variability of XID 403 would be associated with changes of accretion rate. The  Eddington ratio constraints for the six epochs are, as estimated in \hbox{Section \ref{sec3.4}}, $<0.014$, $<0.12$, \hbox{$\approx0.33$}, $<0.018$, $<0.012$, and $\approx1.2$. %The optical/UV emission of \xid\ should not be modified by extinction either.
%The keck/MOSFIRE observation reveals that there was no any broad emission lines in 2012 (between epochs 5 and 6),
%The type 2 classification by \cite{Trump2013} was based on the Keck/MOSFIRE observation in 2012 that is between epochs 5 and 6, 
%and thus \xid\ should not be able to produce any broad emission lines at that time.
%In this state, \xid\ could be considered as a ``true'' type 2 or ``naked'' AGN \citep[e.g.,][]{Netzer2015,Elitzur2016}, and it was accreting with a very low Eddington ratio.
%During its Keck spectroscopic observations in September to October 2012, the Eddington ratio was  low ($<0.012$), and thus it was probably not able to produce any broad emission lines, leading to the type 2 classification in \cite{Trump2013}. In this state, \xid\ could be considered as a ``true'' type 2 or ``naked'' AGN \citep[e.g.,][]{Netzer2015,Elitzur2016} that is accreting with an advection-dominated accretion flow.

However, there are a few properties that are not easily explained by this scenario.
\begin{enumerate}

\item 
The epoch 3 outburst appears to evolve too rapidly for significant changes of the accretion rate, e.g., the flux decreased by a factor of $>6.0$ in $\lesssim6.1$~days. 
Adopting the standard thin-disk model \citep[e.g.,][]{SSD,Netzer2013}, we estimate the radius of the 2500 \AA\ emitting region to be $\approx 900~R_{g}$, where $R_{g} = G\mbh/c^2$ is the gravitational radius. Assuming a disk viscosity parameter of 0.3 and a disk aspect ratio of 0.05, the corresponding thermal and heating/cooling front timescales (Equations 6 and 7 in \citealt{Stern2018}) at this radius are 1 month and 1.5 years, respectively. These results indicate that the accretion rate of XID~403 should not vary significantly within a few days. 

\item In the high state (epoch 6), XID 403 should have produced stronger AGN UV emission than that constrained from the stable $R$-band light curve (comparing the expected AGN SED and the $R$-band upper limit in Figure~\ref{sed}). This is not a critical problem. As introduced in Section~\ref{sec3.4}, mild dust reddening should be able to suppress the $R$-band flux significantly. 

\item It is unusual that from the low state (e.g., \hbox{epoch 5}; $\lambda_{\rm Edd} < 0.012$) to the high state, \xid\ changes from a low accretion rate AGN  to a super-Eddington accreting AGN ($\lambda_{\rm Edd} = 1.2^{+2.9}_{-0.7}$) within rest-frame $\approx1.5$~years. 
However, the substantial uncertainties associated with the $\mbh$ and $L_{\rm bol}$ estimates make the above $\lambda_{\rm Edd}$ constraints rather uncertain.

\end{enumerate}
Overall, we consider the \xray\ variability of \xid\ unlikely to be caused by changes of accretion rate, mainly due to the significant and rapid variability observed between epoch 2--4 as discussed in the first point above.

\subsection{Tidal Disruption Events in an Inactive Galaxy?}\label{sec4.2}
\begin{figure}
    \centering{
    \includegraphics[width=3.3in]{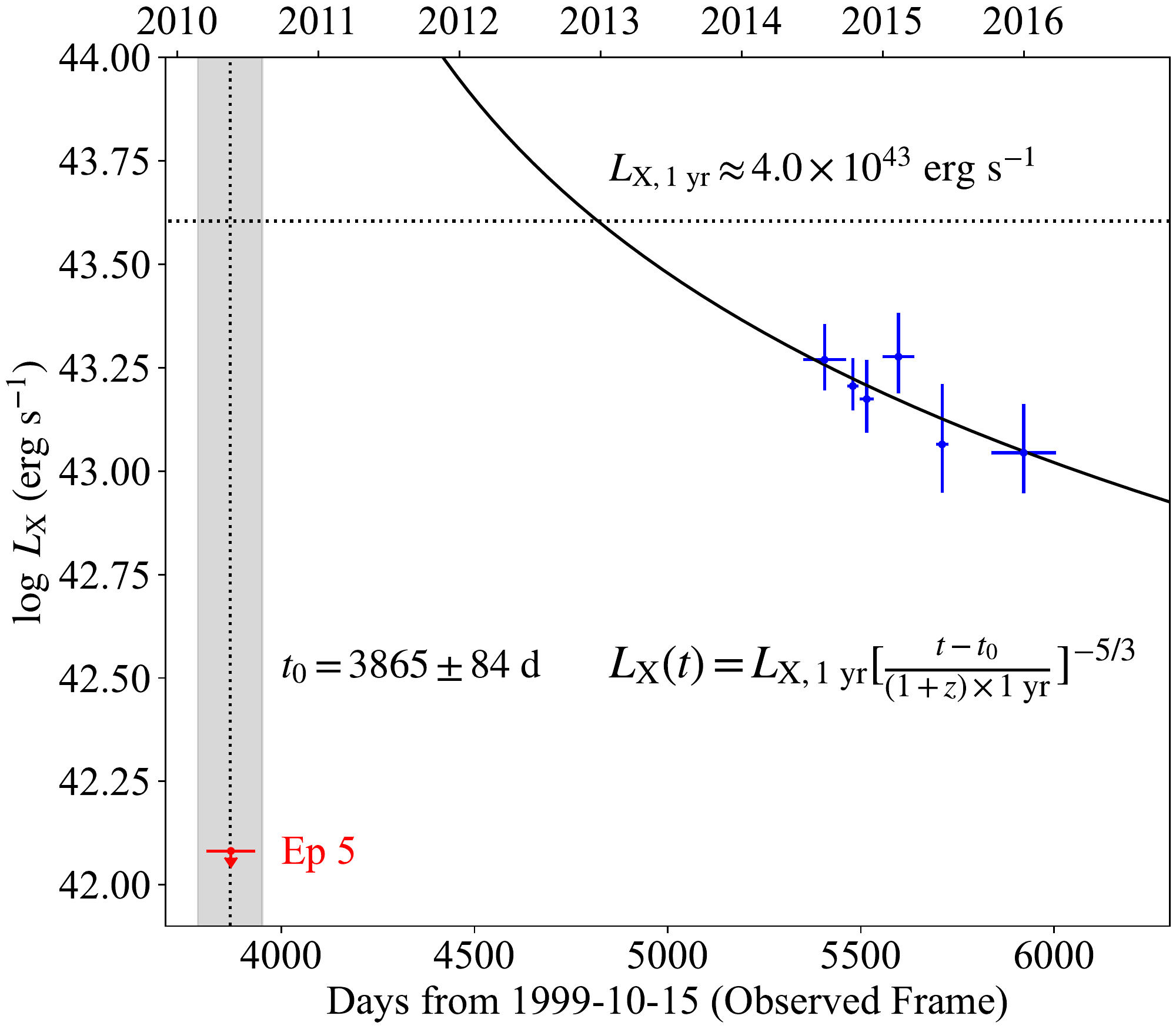}}
    \caption{Epoch 6 X-ray light curve fitted with the canonical $t^{-5/3}$ power law. The red arrow shows the epoch 5 luminosity upper limit.  The vertical dotted line and the shaded region indicate the best-fit $t_0$ and its uncertainties, which overlap epoch 5.  The horizontal dotted line shows the best-fit $L_{\rm X,1 yr}$.}
    \label{tdelc}
\end{figure}

Since typical TDEs are found in inactive galaxies, and there is no clear AGN signature for \xid\ (e.g., Section \ref{prop}), we consider it an inactive galaxy in the TDE scenario.
In this case,  the \xray\ outbursts arise from TDEs.
%variability is due to TDEs, and there is no obscuration of the \xray\ emission. 
There were no contemporary optical/UV outbursts given the \rband\ light curve. This is not unusual for \xray\ TDEs \citep[e.g.,][]{Komossa2015,Saxton2021}. 

To derive physical parameters for the epoch 6~TDE, we fit the epoch 6 light curve with the canonical $t^{-5/3}$ power law \citep[e.g.,][]{Rees1988,Komossa2015,Saxton2021}, expressed as:
\begin{equation}
    L_{\rm X}(t) = L_{\rm X,1~yr}[\frac{t-t_0}{(1+z)\times 1~\rm yr}]^{-5/3}
\end{equation}
where $t_0$ is the time at which the disruption event occurs, and $L_{\rm X,1~yr}$ is the \xray\ luminosity one year later. The \xray\ luminosities for the six subepochs are calculated from their soft-band PFs and the epoch 6 best-fit power-law model.
We use the module {\sc kmpfit} in the Python package {\sc Kapteyn} \citep{Kapteyn} to fit the light curve. The best-fit $t_0$ is $3865\pm84$ days from 1999-10-15, and the best-fit $L_{\rm X,1~yr}$ is $(4.00\pm0.03) \times 10^{43}~\rm erg~s^{-1}$. The parameter uncertainties are estimated from $\Delta \chi^{2}=1$. 
The best-fit TDE light curve is plotted in Figure \ref{tdelc}.  Since the epoch~3 outburst appears transient, we are not able to perform the above analysis for epoch~3.

We note that the best-fit $t_0$ value overlaps the date span of epoch~5, while the epoch~6 TDE should occur after epoch~5. One possibility is that this $t_0$ value is not correct as the epoch~6 light curve does follow exactly the best-fit power-law model (Figure~\ref{tdelc}). This could be either due to large uncertainties of the photometric measurements or our binning scheme of the sparsely sampled epoch~6 data. Another possibility is that the TDE light curve does not necessarily follow the canonical $t^{-5/3}$ power law (e.g., \citealt{Auchettl17}, and references therein).

There are also other unusual aspects regarding the TDE scenario, discussed as follows.

\begin{enumerate}
%\item XID 403 must be a ``dead'' (inactive) AGN. If the SMBH were still actively accreting, we would expect to see strong \xray\ emission in \hbox{epochs 1{--}5}. The inferred 3{--}20 keV \xray\ luminosity is  \hbox{$\approx1.6\times10^{44}~\ergs$} from the [\ion{O}{3}] luminosity \citep{Heckman2005}, which is $\approx 120$ times (adopting a $\Gamma = 2$ spectrum) larger than the upper-limit constraint in epoch 5. Thus, \xid\ must be inactive, and the type 2 classification in \cite{Trump2013} is due to relic narrow line region emission from past activity \citep[e.g.,][]{Merloni2015,Zabludoff2021}.

\item There are two \xray\ outbursts (epochs 3 and 6), and the \xray\ light curve (Figure \ref{xlc}) shows that it is  unlikely to be caused by a single TDE plus fluctuations.
It is very unlikely to have two TDEs given the short time interval of $\approx$ 2.5 years and the typical \xray\ TDE rate of $\approx10^{-4}\textrm{--}10^{-5}~\rm yr^{-1}~galaxy^{-1}$ \citep[e.g.,][]{Donley2002,Komossa2015,Sazonov2021}.
%It is not impossible to have two TDEs in rest-frame $\approx$ 2.5 years, but the chance should be slim considering the typical TDE rate of  $\approx10^{-4}\textrm{--}10^{-5}~\rm yr^{-1}~galaxy^{-1}$. 
There have also been suggestions that partial TDEs \citep[e.g.,][]{Chen2021a,Payne2021,Chen2022,Wevers2022} or TDEs in SMBH binaries \citep[e.g.,][]{Liu2009,Shu2020} might produce multiple TDE flares. However, there are only a few such candidates, and their \xray\  light curves differ from that of \xid.
%\xray\ light curves of the limited examples of such candidates are not similar to the light curve of \xid.
%, but these models do not predict \xray\ light curves similar to that of \xid.
Recently, a few \xray\ quasi-periodic eruptions (QPEs) have been reported \citep[e.g.,][]{Miniutti2019,Arcodia2021} in both active and inactive galaxies. We are not able to detect any periodicity in the \hbox{epoch 6} light curve, but the shapes of the \xray\ light curves and the overall properties of these QPE objects also do not resemble those of \xid.

\item 
%The epoch 6 spectrum shows a relatively flat 
%power law ($\Gamma = 2.8\pm0.3$) compared to
The epoch 6 spectral shape (a $\Gamma = 2.8\pm0.3$  power law  in the 1.3{--}13 keV band) is  not consistent with typical TDEs.
The \xray\ spectra of  typical \xray\ TDEs can be described with  single black-body (temperatures of $\approx 10\textrm{--}100~\rm eV$) or steep power-law ($\Gamma \gtrsim 4$; typically not extending to $\gtrsim 5$ keV energies) models \citep[e.g.,][]{Komossa2015,Saxton2021}.
%with a single black body with a temperature of  $\approx 10\textrm{--}100~\rm eV$, or with a much steeper power law with $\Gamma > 4$ \citep[e.g.,][]{Komossa2015,Saxton2021}. 
For the black-body modeling of the epoch 6 spectrum, the best-fit temperatures are much higher ($\gtrsim 6$ times) than  typical TDE temperatures (see Section \ref{sec3.2} and Table~\ref{tab3}). 
However, there are indeed a couple unusual TDE candidates that show similar spectral shapes. One example is XMMSL2 J144605.0$+$685735 reported by \cite{Saxton2019}. At a redshift of 0.029, its \hbox{0.3{--}10 keV} \xray\ spectrum can be described by a $\Gamma\approx2.6$ power law. Another object is 3XMM J150052.0$+$015452 at a redshift of 0.145 \citep{Lin2017,Lin2022,Cao2023}. Its \hbox{0.3{--}10~keV} \xray\ spectra observed after 2015 can be described with a black-body component with a temperature of $kT_{\rm diskbb}\approx\rm 0.15~keV$  plus $\Gamma=2.5$ power law. Besides the epoch 6 spectrum, the flat spectral shape ($\Gamma=1.2^{+0.7}_{-0.6}$) of the epoch~3 spectrum is also unusual, although the $\Gamma$ uncertainties are large.

%\item
%\textbf{The best-fit $t_0$ is within epoch 5 (red upper limit in Figure \ref{tdelc}), but TDE should occurred after the end time of epoch~5. It could be due to the 6 data points are close in time ($\approx251$ rest-frame days), and the light curve does not show significant variability (Section \ref{sec:LC}). Thus the $t_0$ value as well as $L_{\rm X,1~yr}$ are poorly constrained, although the uncertainties are relatively small. Another possibility is that the TDE light curve does not strictly follow the canonical $t^{-5/3}$ law. Other power-law indice are also suggested by theoretical models (e.g., \citealt{Auchettl17}, and reference therein).
%The decays of the two slow TDE candidates (XMMSL2 J144605.0$+$685735 and 3XMM J150052.0$+$015452) are both very slow. \xid\ could be also a slow TDE candidate given that there is no significant variability within rest-frame $\approx251$ rest-frame days.}

\end{enumerate}

We do not find any TDE candidates that resemble XID~403, mainly due to the two outbursts from XID~403. Besides XMMSL2 J144605.0$+$685735 and 3XMM J150052.0$+$015452 that share some similarity in the \xray\ spectral shapes, another unusual TDE candidate that is probably relevant is 1ES 1927$+$654 \citep[e.g.,][]{Trakhtenbrot2019,Ricci2020,Ricci2021,Laha2022,Masterson2022}. The optical light curve of this source shows an outburst followed by a TDE-like fading \citep[e.g.,][]{Trakhtenbrot2019}. The \xray\ light curve shows a strong dip, and then a constant increase in luminosity to levels exceeding the pre-outburst level. During the \xray\ rise, a very steep power-law component ($\Gamma\approx 3$) appears. This is interpreted as the destruction and re-creation of the inner accretion disk and corona, perhaps by interactions between the accretion disk and debris from a
tidally disrupted star \citep[e.g.,][]{Ricci2020}.  Although the  epoch 6 \xray\ spectral properties of \xid\ are similar to those of 1ES 1927$+$654 in 2018 and 2019, the optical and \xray\ light curves  differ significantly from those of 1ES 1927+654.

 Overall, we cannot exclude the TDE scenario for explaining the \xray\ outbursts of \xid. But in this case, it should be a TDE candidate with unusual properties (two outbursts, unusual spectral shapes), and it would be the highest-redshift TDE candidate detected so far.  
 %we do not favor the TDE scenario for explaining the \xray\ variability and multiwavelength properties of \xid.

\subsection{Change of  Obscuration?}\label{sec4.3}
\begin{figure}
\includegraphics[width=3.3in]{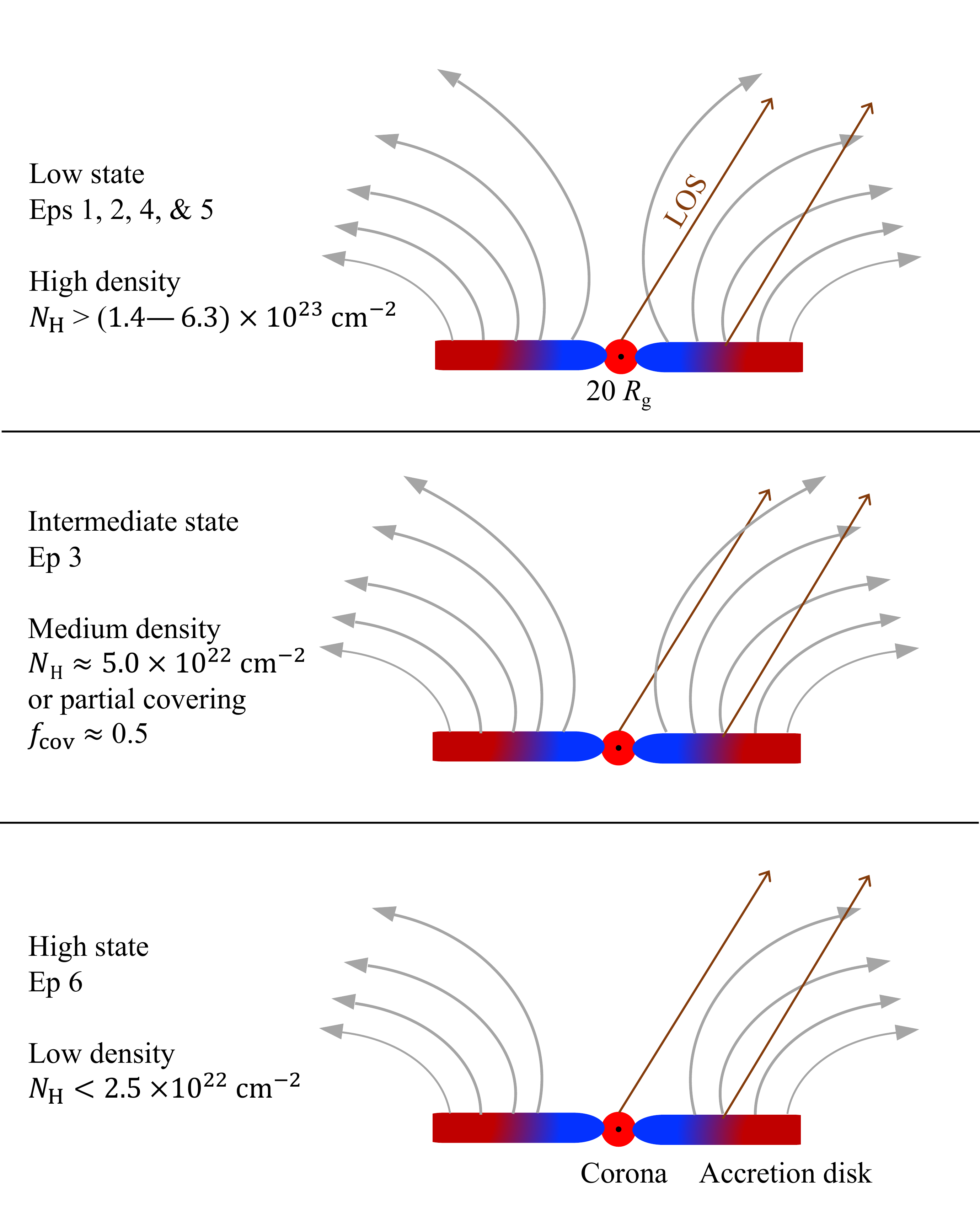}
\caption{Schematic diagram showing the \xray\ unveiling events in \xid\ (not to scale). 
%The blue region represents the clumpy dusty torus with the clumps denoted by the blue dots. In the three panels, \hbox{X-rays} from the corona were absorbed by high-density (\hbox{$> 3.4\textrm{--}6.1 \times 10^{23}~\rm cm^{-2}$}), medium density ($\approx 1.3 \times 10^{23}~\rm cm^{-2}$), and low-density ($< 2.5~\rm cm^{-2}$) regions of the torus along the line of sight, respectively. The changes of the absorption column densities are probably due to the motion of the torus clumps.
The gray curves represent the \xray\ absorber that is variable in covering factor and column density. The lines of sight to the corona and the accretion disk are indicated by the brown arrows.
}
\label{sce}
\end{figure}

%In the scenario of change of dust-free gas obscuration, \xid\ is a NLS1 with a high accretion rate ($\lambda_{\rm Edd} \approx 1.2$). 
In the scenario of change of obscuration, \xid\ is a luminous AGN with intrinsic \xray\ emission described by the epoch 6 data. It accretes at a high  rate ($\lambda_{\rm Edd} = 1.2^{+2.9}_{-0.7}$), and it has a steep power-law \xray\ spectrum with $\Gamma = 2.8\pm 0.3$. The AGN SED and the coronal \xray\ emission do not vary, and  the observed \xray\ variability  arises from changes of the line-of-sight obscuration, which could be due to
changes of either the column density or the covering factor
($f_{\rm cov}$) of the absorber.
%The change could be due to changes of  column density. And another possible explanation is changes of partial-covering absorption, where the \nh\ value does not vary but a fraction of coronal emission was able to leak through the absorber.
In epochs 1, 2, 4, and 5, the \xray\ corona was fully covered ($f_{\rm cov}=1$) by the absorber with high column densities. 
In epoch~3, the column density or the covering 
factor ($f_{\rm cov}<1$) was lower. 
In epoch~6, the column density was the lowest and the line of sight to the \xray\ corona was largely cleared.
In this scenario, the \xray\ outbursts in epochs 3 and 6 are the consequence of two \xray\ unveiling events. A schematic illustration of this scenario is shown in Figure \ref{sce}.

For epoch 6, the column density is constrained to be \hbox{$< 2.5 \times 10^{22}~\rm cm^{-2}$} (Section \ref{sec3.2}). We then estimate the absorber properties for the other epochs, adopting the epoch~6 spectrum as the 
intrinsic coronal emission.
%column density constraints for epochs 1{--}5. 
For epoch 3, we first fit the spectrum 
using an absorbed power-law model (phabs*zphabs*zpowerlw)
with $\Gamma$ and power-law normalization fixed at
the epoch 6 values. The resulting intrinsic 
absorption column density is $\approx 5.0 \times 10^{22}~\rm cm^{-2}$ (Table~\ref{tab3}). We then fit the spectrum
with a simple power-law model (phabs*zpowerlw), fixing
only the $\Gamma$ value. The resulting 
power-law normalization (Table~\ref{tab3}) is $\approx50\%$ of the epoch~6 value, indicating that the spectrum could be
alternatively described by partial-covering absorption
with a high $N_{\rm H}$ and $f_{\rm cov}\approx0.5$.

For each of the other four epochs, we derived an $N_{\rm H}$ lower limit that can reduce the photon fluxes to the upper limit values adopting the best-fit spectrum in epoch 6 as the intrinsic spectrum.  The resulting $N_{\rm H}$ constraints for epochs 1, 2, 4, and 5 are $> 5.6$, $> 1.4$, $> 5.0$, and $> 6.3\times10^{23}~\rm cm^{-2}$, respectively.

Change of obscuration could occur in both type~1 and type~2 AGNs (Section~\ref{sec:intro}), but the locations of the absorbers might differ (i.e., in the torus, BLR, or disk wind). 
From the variability time between epoch 2--4, we estimate the distance ($D$) of the absorber to the \xray\ corona. 
The rise time and decay time of the epoch~3 outburst are both  $\lesssim6.1$ rest-frame days.
%From epoch 3 to epoch 4, the  soft-band photon flux decreased by a factor of \hbox{$>6.0$} in $\approx6.1$ rest-frame days.
We adopt a corona size of $20~R_{g}$ \citep[e.g.,][]{Chartas2002,Chartas2009,Risaliti2009}. The absorber should move a distance larger than 
this size in 6.1 days, and thus the velocity ($v$) should be larger than  $20~R_{g}/t$, where $t = 6.1$ days.
 Assuming that the absorber is moving with a Keplerian velocity, we obtain:
\begin{equation}
D = \frac{G\mbh}{v^2} <\frac {G\mbh t^2}{(20R_{g})^2} \approx 324~\rm light\textrm{-}days.
\end{equation}
We then compare the distance constraint to the  BLR radius and inner torus radius estimated from empirical radius-luminosity relations \citep[e.g.,][]{Bentz2013,Minezaki2019}. Using the \cite{Bentz2013} relation, we estimate an H$\beta$ BLR radius of $\approx 30$ light-days from the \hbox{5100 \AA} luminosity of the inferred AGN SED in Figure \ref{sed}. Using the \cite{Minezaki2019} relation, we estimate a dust-sublimation radius (the lower boundary on the torus radius)
of  $\sim 135$ light-days from the $V$-band
luminosity of the inferred AGN SED.
The upper limit on the absorber location ($<324$ light-days) is close to the inner radius of the torus. 
Since the torus has an extended structure, it is likely that the absorber is located in a smaller-scale region, containing dust-free gas. Also considering that we are observing the intrinsic \xray\ emission in epoch 6, \xid\ is likely not affected by torus obscuration, and it  should be a type~1 AGN.

Luminous \xray\ emission lasted over the entire \hbox{epoch 6}, i.e., 251 rest-frame days; we use this time to constrain the size ($l$) of the low-density region ($< 2.5 \times 10^{22}~\rm cm^{-2}$). Assuming that the absorber  moves at the same velocity as that constrained from the epoch 3,  we obtain:
\begin{equation}
l \geq  251 \times \frac{20~R_{g}}{6.1} = 1915~R_{g} \approx 0.54~\rm light\textrm{-}days.
\end {equation}
Between epoch 5 and epoch 6, the absorber column density changed by more than
an order of magnitude. Given these constraints, the absorber is again unlikely the torus as 
it is probably difficult to support
dynamically such a large size ($l~\times$ thickness of the torus) low-density ``hole'' in the torus.  
%The absorber is probably not composed of both  
%low-density and high-density regions ($> 6.3 \times 10^{23}~\rm cm^{-2}$), as it might be difficult to support
%dynamically a large size low-density ``hole''. 
Instead, the change of the
column density might be intrinsic to the absorber (e.g., a variable disk wind)
or simply be due to a reduction of the absorber covering factor that exposes
the \xray\ corona (e.g., Figure \ref{sce}). 

Therefore, the unusual \xray\ variability of \xid\ can be consistently explained by changes of obscuration from a small-scale dust-free absorber in a type 1  AGN. Such an absorber should not attenuate the accretion-disk continuum emission or the BLR line emission. However, there is no  AGN signature  in the SED or spectra. 
%and thus we would expect AGN signatures in the SED or spectra. 
One likely explanation is host-galaxy dilution plus dust extinction, which we explore in the following points.

\begin{enumerate}

\item  The observed FIR--UV SED appears to be dominated by a star-forming host galaxy (Section \ref{secSED}). The expected AGN SED based on the epoch 6 \xray\ spectrum would
produce an $R$-band flux that is $\approx1.5$ times brighter than the observed value. As introduced in Section~\ref{sec4.3}, one possibility is that  the AGN UV emission is affected by dust extinction from the host galaxy (due to the intense star formation) or AGN polar dust \citep[e.g.,][]{Gilli2014,Lopez-Gonzaga2016,Asmus2019,Buat2021}.
For example, an $E(B-V)$ value of 0.24 mag would be able to suppress the \rband\ flux by a factor of 4.6 (1.5/0.33), adopting a Small Magellanic Cloud extinction curve \citep{Gordon2003}. The corresponding $N_{\rm H}$ value is only $1.4 \times 10^{21}~\rm cm^{-2}$ assuming a gas-to-dust ratio of
$\nh/E(B-V)=5.9\times10^{21}~\rm mag^{-1}~cm^{-2}$ \citep{Rachford2009}, still within the upper-limit constraint from the epoch 6 \xray\ spectrum. Thus, it is possible to observe a host-dominated SED due to  dust extinction.

\item 
There was no broad H$\alpha$ emission line in the Keck/MOSFIRE spectrum (Section \ref{sec:data}). It is not unusual to find \xray\ unobscured AGNs in \xray\ surveys
that do not show emission-line signatures in their optical spectra 
(i.e., optically ``dull''
AGNs), and most of them are explained by host-galaxy dilution, with dust
extinction being another possible mechanism \citep[e.g.,][]{Comastri2002,Severgnini2003,Trump2011,Merloni2014,Fitriana2022}. Adopting the 
{\sc cigale} best-fit galaxy spectrum (Section \ref{secSED}) 
and the \cite{Vanden2001} mean quasar spectrum scaled to the expected AGN SED (with allowed scatter) in Figure \ref{sed},
we simulate a set of Keck/MOSFIRE spectra considering the spectral signal-to-noise ratio (SNR) at
each wavelength.
We visually inspect these spectra and a significant fraction ($\approx40\%$) of them
do not show a clear broad H$\alpha$ emission line. 
Thus it is possible that XID 403, as a type 1 AGN, does not exhibit a 
broad H$\alpha$ emission line in the Keck/MOSFIRE spectrum.

\end{enumerate}

\subsubsection{A High-Redshift analog of NLS1s and Connections}

%\begin{figure*}
%\centerline{
%\includegraphics[width=6.0in]{fig6.pdf}}
%\caption{Similar with Figure \ref{sed}, but $\Delta\alpha_{\rm OX}$ of the AGN SED is 0.165. The gray curve represents the galaxy SED, which is the 
%{\sc cigale} best-fit SED for the AGN-subtracted data. The black curve is the sum of galaxy SED and AGN SED.
%}
%\label{fig6}
%\end{figure*}

Overall, the change-of-obscuration scenario appears to explain naturally the \xray\ variability and multiwavelength
properties of \xid, where the 
observed \xray\ emission is
modified by various amounts of absorption from a small-scale dust-free
absorber and the optical/UV emission is dominated by the host galaxy.
Since \xid\ is probably accreting at a super-Eddington accretion rate
given the large estimated $\lambda_{\rm Edd}$ value  ($1.2^{+2.9}_{-0.7}$) and the large 
epoch 6 \xray\ photon index ($2.8\pm0.3$), 
it appears to be a high-redshift analog of 
local NLS1s that are considered to have super-Eddington accretion rates and 
sometimes exhibit strong and rapid \xray\ variability (e.g., Mrk 335, \citealt{Gallo2018}; NGC 4051, \citealt{Guainazzi1998};  1H 0707$-$495, \citealt{Fabian2012}; IRAS 13224$-$3809, \citealt{Boller1997,Fabian2013}).\footnote{Considering the substantial uncertainties associated with $\mbh$ and $\lambda_{\rm Edd}$, we cannot exclude the possibility that XID~403 has a moderate Eddington ratio, and then it would belong to a new type of extremely X-ray variable AGNs with unusually steep X-ray spectra. However, Occam's razor would favor the simpler NLS1 analog explanation.}

Recently, an increasing number of type 1 quasars have been found to display
similar strong \xray\ variability (e.g.,  PHL 1092, \citealt{Miniutti2012}; SDSS J0751$+$2914, \citealt{Liu2019}; SDSS J1539+3954, \citealt{Ni2020};  PG 1448+273, \citealt{Laurenti2021}; SDSS J1350$+$2618, \citealt{Liu2022}). These quasars show also
signatures of high or even super-Eddington accretion rates, e.g., 
large \xray\ photon indices, large estimated $\lambda_{\rm Edd}$ values, and/or
weak [\ion{O}{3}] emission 
(being part of the Eigenvector 1 parameter space; e.g., \citealt{Boroson1992,Sulentic2000,Shen2014}), and their \xray\ variability
likely shares the same origin as that in NLS1s.
A common feature of the \xray\ variability in these AGNs is the lack of
contemporaneous UV/optical continuum or emission-line variability, 
indicative of a stable accretion rate and little UV extinction
from the \xray\ absorber.
Though in the case of XID 403, it is not feasible to assess the
UV/optical variability due to the
strong host-galaxy dilution.
Another important characteristic of these AGNs
is that they follow the $\alpha_{\rm OX}\textrm{--}L_{2500~\textup{\AA}}$ 
relation in the highest \xray\ flux states and they become \xray\ weaker
in the low states
\citep[e.g.,][]{Miniutti2012,Liu2019,Liu2021,Ni2020,Boller2021},
which also supports the obscuration scenario where the highest
state represents the unobscured intrinsic coronal emission.

A good candidate for the \xray\ absorber in \xid\ is a powerful
accretion-disk wind that is generally expected in AGNs with high accretion
rates \citep[e.g.,][]{Takeuchi2014,Jiang2019,Giustini2019}.
The dynamical nature of the wind causes variable obscuration, and
the high-density clouds in the wind 
produce heavy or even Compton-thick obscuration
in the low state.
As illustrated in \hbox{Figure \ref{sce}}, the corona is shielded heavily by the wind
in the low states (epochs 1, 2, 4, and 5).
In the epoch 3 intermediate state, the wind weakened with a small $N_{\rm H}$ or a less than 100\% covering factor of the corona, allowing some of the  \xray\ photons to pass through. In epoch 6, the wind did not intercept the 
line of sight to the corona, and the intrinsic \xray\ emission was observed.
This scenario is consistent with the thick disk and outflow (TDO) model
recently proposed to explain the strong \xray\ variability in weak emission-line
quasars and other similar AGNs with high accretion rates \citep{Ni2020,Ni2022}.
The small-scale TDO is probably clumpy, and it may sometimes shield the corona
partially, yielding a steep \xray\ spectrum (with little 
absorption signature) dominated by 
the leaked fraction of the intrinsic continuum \citep[e.g.,][]{Liu2019,Liu2022,Wangcj2022}.

The fraction of super-Eddington accreting AGNs that displays
strong \xray\ variability is poorly constrained ($\sim15\%$; \citealt{Liu2019}),
and it is likely related to the inclination angle
of the system
and the covering factor of the wind, which also affect the duration or duty 
cycle of the \xray\ dimming event. \xid\ was discovered in the 7~Ms
\hbox{CDF-S} \xray\ survey, and there were no other type 1 AGNs in this survey reported to show such strong
\xray\ variability \citep{Yang2016,Zheng2017}.
%$and it was the only type 1 AGN in the survey 
%$that shows such strong
%\xray\ variability \citep{Yang2016,Falocco2017,Zheng2017}.
Most of the \hbox{CDF-S} AGNs are obscured type 2 AGNs. 
From the \cite{Luo2017} CDF-S \xray\ source catalog, 
we find only 10 AGNs that have effective photon indices of $>2$
(unobscured and probably having high accretion rates) 
and more than 100 net counts
in the 0.5--7 keV band (for variability assessment). Thus 
it is not surprising that only \xid\ (one out of 10) 
was found to be 
affected by variable wind obscuration.\footnote{Another AGN among these 10 sources, XID 479,
was reported in Appendix C of \cite{Yang2016} to have a luminosity variability amplitude
of \hbox{$\approx 2$} with little $N_{\rm H}$ variability (almost always unobscured), and it 
was also variable in the optical.
Thus the variability is probably due to changes of accretion rate.}

X-ray obscuration events have also been observed in a small number of more typical type 1
AGNs, e.g., NGC 5548 \citep[e.g.,][]{Cappi2016} and NGC 3227 \citep[e.g.,][]{Mao2022,Wang2022}. These
events generally have shorter durations (sometimes being transient) and lower variability amplitudes (with \hbox{$N_{\rm H}\approx10^{22}\textrm{--}10^{23}~\rm cm^{-2}$} in the low
states) than those in super-Eddington accreting 
AGNs. The fraction of AGNs showing such obscuration 
and the frequency of observing such an event \citep[e.g.,][]{Pu2020,Timlin2020} are
also lower than those for super-Eddington accreting AGNs. These obscuration
events are considered to originate from moving clouds in the BLRs,
which are probably also responsible for the absorption variability
in type 2 AGNs in general \citep[e.g.,][]{Risaliti2011,Rivers2015b,Ricci2016,Hickox2018}.
Given the different characteristics, we consider the absorbers in \xid\ and other super-Eddington accreting AGNs different from the 
typical BLR clouds. Although the accretion-disk wind is closely connected 
to the BLR and may even be an essential component of the BLR, the obscuring
clouds in \xid\ should have higher densities 
and be more persistent (e.g., larger
covering factors) than typical BLR clouds.

Besides absorption, another mechanism that is frequently invoked to explain
the variable \xray\ emission in NLS1s is relativistic disk reflection \citep[e.g.,][]{Grupe2008,Fabian2012,Miniutti2012,Parker2014,Grupe2019}.
This scenario \citep[e.g.,][]{Ross2005,Fabian2010} was proposed mainly to describe the 
reflected continuum, broad iron line, and soft excess in the low \xray\ 
state. Changes of the corona height in the
lamppost geometry affect the observed \xray\ emission, but
these are generally not used to account for the entire 
\xray\ variability of such AGNs.
Significant normalization changes in the coronal emission (i.e., changes of
accretion rates) are often required when describing multi-epoch 
spectra using solely the disk-reflection model \citep[e.g.,][]{Miniutti2012,Waddell2019}; otherwise, 
variable absorption is needed in addition to the disk reflection 
\citep[e.g.,][]{Boller2021}. The available low-state spectrum (epoch
3)
of XID 403 has limited photon statistics, and thus we are not able to
assess if disk reflection is present. Nevertheless, it may contribute to some extent to the observed \xray\ variability.

The unprecedented deep CDF-S data have thus likely revealed, for the first time,
a high-redshift moderate-luminosity
AGN that shows strong and rapid \xray\ variability similar to local
NLS1s. This is consistent with our current understanding that the
basic emission properties of the AGN population do not have any significant 
redshift evolution. Future deep \xray\ surveys might be able to uncover 
a sample of such objects for detailed studies.

 \section{Summary and Future Work}\label{sec:summary}
In this paper, we report  the   extreme \xray\ variability of XID 403 in the 7 Ms CDF-S. The optical counterpart of \xid\ has a redshift of $z = 1.608$. Our results are summarized as follows.
\begin{enumerate}
\item In epochs 1, 2, 4, and 5, \xid\ was in a low state, and it was not detected by \chandra.
 In epoch~3, \xid\ brightened to an intermediate state.
 The full-band (0.5{--}5~keV)  PF increased by a factor of \hbox{$>2.5$}   within \hbox{6.1
 days}. The outburst lasted for \hbox{$\approx5.0$--7.3 days},   and then the full-band PF decreased by a factor of $>6.0$ within 6.1 days. The 0.5{--}5~keV (rest-frame \hbox{1.3{--}13 keV}) \xray\ spectral shape appears flat ($\Gamma=1.2^{+0.7}_{-0.6}$).
 In epoch 6, \xid\ entered a high state, lasting  $>251$ days. Compared to epoch 5, the PF increased by factors of $> 12.6$, $> 12.1$, $> 3.1$ in the full, soft, and hard bands, respectively. There is no apparent flux or spectral-shape variability within \hbox{epoch 6}. The 0.5{--}5~keV \xray\ spectrum can be described by a simple power-law model ($\Gamma = 2.8\pm0.3$) modified  by the Galactic absorption.
 See Sections \ref{sec:LC} and \ref{sec3.2}.

\item 
There is no significant optical/UV variability and there is no $R$-band brightening contemporaneous with the \xray\ outburst. The observed SED is dominated by the host galaxy. The stellar mass and the SFR from the {\sc cigale} best-fit results are $(2.7\pm0.2)\times10^{10}~\msun$ and $(46\pm3)~\msun~\rm yr^{-1}$.     See Sections \ref{sec2.2} and \ref{secSED}. 

\item Among the three scenarios discussed in \hbox{Section \ref{sec:discussion}}, we prefer the change-of-obscuration scenario \hbox{(Section \ref{sec4.3})},  as it explains most naturally the \xray\ variability and multiwavelength properties of \xid. 
The \xray\ variability is due to two \xray\ unveiling events, where the line of sight to the corona is no longer shielded by high column-density gas clumps in a small-scale dust-free absorber. \xid\ is likely a high-redshift analog of NLS1s, 
and the absorber is probably a powerful accretion-disk wind  driven by super-Eddington accretion.
We cannot exclude the possibility that XID 403 is an unusual TDE candidate (Section~\ref{sec4.2}).

\end{enumerate}

The discovery of XID 403 involved unique data sets, and it is the highest-redshift moderate-luminosity AGN discovered with such extreme \xray\ variability. For future prospects, a high SNR NIR spectrum of XID~403, e.g., with   
the Near InfraRed Spectrograph (NIRSpec) of the  James Webb Space Telescope ({\it JWST}; \citealt{Gardner2006,Jakobsen2022}),
might be able to reveal broad Balmer emission lines and thereby nail down its type 1 nature.
\xid\ is in the Deep-Drilling Fields (DDFs) of the Vera C. Rubin Observatory (e.g., \citealt{Ivezic2008,Brandt2018}), and it will soon have superb photometric monitoring for $\approx10$ years. 
In our preferred scenario, change of obscuration, XID 403 is a high-redshift analog of NLS1s with strong and rapid \xray\ variability. 
It might be feasible to search
the \chandra\ archive for regions covered by multiple
$\gtrsim100$~ks exposures and identify
high-redshift 
moderate-luminosity AGNs with similar X-ray 
variability.
The extended Roentgen Survey with an Imaging Telescope
Array ({\it eROSITA}; \citealt{Merloni2012,Predehl2017}),
if it resumes operation, will be able to discover more NLS1s with such \xray\ variability. The next generation \xray\ instruments like the new Advanced Telescope for High-ENergy Astrophysics ({\it Athena}; \citealt{Barcons2017})
, the Survey and Time-domain Astrophysical Research eXplorer ({\it STAR-X}; \citealt{Zhang2017}), and the Advanced X-ray Imaging Satellite ({\it AXIS}; \citealt{Mushotzky2019,Marchesi2020}) may be able to reveal a population of objects like \xid\ at high redshifts via deep surveys.

\acknowledgments
We thank the anonymous referee for helpful comments and suggestions.
L.Y. and B.L. acknowledge financial support from
the National Natural Science Foundation of China
grant 11991053,
China Manned Space Project grants NO. CMS-CSST-2021-A05
and NO. CMS-CSST-2021-A06.
W.N.B. acknowledges support from the V.M. Willaman Endowment and CXC grant AR1-22006X.
FEB acknowledges support from ANID-Chile BASAL AFB-170002, ACE210002, and FB210003, FONDECYT Regular 1200495 and 1190818, 
and Millennium Science Initiative Program  – ICN12\_009.  
Y.Q.X. acknowledges support from NSFC-12025303 and 11890693, the K.C. Wong Education Foundation, and the science research grants from the China Manned Space Project with NO. CMS-CSST-2021-A06.

\bibliographystyle{aasjournal}
\bibliography{ref}

\end{document}